%% file: lrm07.tex
\documentstyle[epsf]{article}
\title{Thermodynamic properties
       \protect\\
       of the periodic nonuniform spin-$\frac{1}{2}$ isotropic $XY$ chains
       \protect\\
       in a transverse field}
\author{Oleg Derzhko$^{\dagger,\ddagger}$,
	Johannes Richter$^{\star}$
        and
        Oles' Zaburannyi$^{\dagger}$\\
\small   {\em $^{\dagger}${Institute for Condensed Matter Physics}}\\
\small   {\em {1 Svientsitskii St., L'viv-11, 290011, Ukraine}}\\
\small   {\em $^{\ddagger}${Chair of Theoretical Physics,
                            Ivan Franko State University of L'viv}}\\
\small   {\em {12 Drahomanov St., L'viv-5, 290005, Ukraine}}\\
\small   {\em $^{\star}${Institut f\"{u}r Theoretische Physik,
			 Universit\"{a}t Magdeburg}}\\
\small   {\em {P.O. Box 4120, D-39016 Magdeburg, Germany}}}

\date{\today}

\begin{document}

\maketitle

\begin{abstract}
Using the Jordan-Wigner transformation and the continued-fraction method
we calculate exactly  the density of states
and thermodynamic quantities
of the periodic nonuniform
spin-$\frac{1}{2}$ isotropic $XY$ chain
in a transverse field.
We discuss in detail the changes
in the behaviour of the thermodynamic quantities
caused by regular nonuniformity.
The exact consideration of thermodynamics is extended
including a random Lorentzian transverse field.
The presented results are used to  study
the Peierls instability
in a quantum spin chain.
In particular, we examine the influence of a non-random/random field 
on the spin-Peierls instability with respect to dimerization.
\end{abstract}

\vspace{1cm}


\vspace{1cm}


\vspace{1mm}

\noindent
{\bf Postal addresses:}\\

\vspace{2mm}

\noindent
{\em
Dr. Oleg Derzhko (corresponding author)\\
Oles' Zaburannyi\\
Institute for Condensed Matter Physics\\
1 Svientsitskii St., L'viv-11, 290011, Ukraine\\
Tel: (0322) 42 74 39\\
Fax: (0322) 76 19 78\\
E-mail: derzhko@icmp.lviv.ua\\

\vspace{1mm}

\noindent
Prof. Johannes Richter\\
Institut f\"{u}r Theoretische Physik, Universit\"{a}t Magdeburg\\
P.O. Box 4120, D-39016 Magdeburg, Germany\\
Tel: (0049) 391 671 8841\\
Fax: (0049) 391 671 1217\\
E-mail: Johannes.Richter@Physik.Uni-Magdeburg.DE

\clearpage

\renewcommand\baselinestretch {1.25}
\large\normalsize

\input{lra07.tex}

\input{lrb07.tex}

\end{document}

%% file: lra07.tex
\section{Introduction}

The study of regularly nonuniform spin models
is an attracting problem of statistical mechanics.
Besides of its general academic importance
the development of magnetic materials in recent years
makes the study of nonuniform spin models particularly interesting  
for experimental application.
In order to achieve a progress in understanding
the generic features generated by periodic nonuniformity
it is desirable to examine simple models
that may be investigated without making any approximation.
Among possible candidates of such systems one can mention
the spin-$\frac{1}{2}$ $XY$ model in one dimension.
The uniform one-dimensional spin-$\frac{1}{2}$ $XY$ model 
in a transverse field 
was introduced by Lieb, Schultz and Mattis.$^{1}$
These authors noticed
that a lot of statistical mechanics calculations
for such a spin model can be performed exactly
since
it can be rewritten
as a model of noninteracting spinless fermions by means
of the Jordan-Wigner transformation.
The nonuniform version of the 
transverse spin-$\frac{1}{2}$ $XY$ chain
also can be mapped onto a chain of free spinless fermions,
however, with an on-site energy and hopping integrals
that vary from site to site. 
Especially attractive is the case of isotropic 
spin coupling
with regularly alternating exchange integrals and transverse fields
since after fermionization one faces a model for which a lot
of work has been done.
One should mention here the results for the tight-binding Hamiltonian
of periodically modulated chains$^{2,3}$
and the
spinless Falicov-Kimball
chain.$^{4,5}$

One of the goals of the present paper
is to give a magnetic interpretation of
those results derived for
the one-dimensional tight-binding spinless fermions.
Exploiting the
continued-fraction
approach developed in the mentioned papers$^{2-5}$
we shall be able to calculate exactly the one-fermion Green functions
and therefore to obtain
the thermodynamic quantities for the periodic nonuniform
spin-$\frac{1}{2}$ isotropic $XY$ chain in a transverse field.
We shall treat few examples of
the periodic nonuniform spin-$\frac{1}{2}$ isotropic $XY$ chain
in a transverse field
in order to reveal
the changes in the thermodynamic properties
induced by periodic nonuniformity.
The model of the considered regularly nonuniform magnetic chain allows
even a natural extension to include additional disorder remaining the model
exactly solvable.
Namely, one can assume the transverse fields to be random
independent Lorentzian variables
with regularly alternating mean values and widths of distribution.
To derive exactly the random-averaged density of states for such a model
one should at first average a set of equations for
the Green functions using contour integrals.$^{6-11}$
As a result
one comes to a set of equations
similar to that for the periodic nonuniform non-random case.

It should be noted that the periodic nonuniform
spin-$\frac{1}{2}$
isotropic $XY$ chain was considered
in several papers$^{12-25}$
dealing mainly with 
the adiabatic treatment of 
the spin-Peierls instability.
However, those papers were focussed
mostly on the influence of the structural degrees of freedom
upon the magnetic ones,
rather than on
the exhaustive analysis of
the properties of a magnetic chain
with regularly alternating exchange couplings.
Another closely related study concerns the
spin-$\frac{1}{2}$ isotropic $XY$ model
on a one-dimensional superlattice$^{26}$.
The treatment reported in Ref. 26,
however, was restricted to the magnon spectrum.
A related study of
the spin-$\frac{1}{2}$ isotropic $XY$ chain in a transverse
field with two kinds of coupling constant aimed on examining the
condition for appearance of an energy gap was reported in Ref. 27.
Finally, let us note that the periodic nonuniform chain can be viewed
as the uniform chain with a crystalline unit cell containing
several sites of the initial lattice
(as a matter of fact such a point of view
was adopted, for example, in Refs. 12, 13)
and thus
the standard methods elaborated for such complex crystals may be
exploited.
However, we prefer to treat periodically nonuniform chains
since from such a viewpoint
an elegant continued-fraction approach
immediately arises
that seems to be a natural and
convenient language for describing such compounds.

The present paper is a more extensive version of the results
briefly reported in Refs. 28, 29
containing more details on the calculation and more applications.
We show that the presented below 
method based on continued-fraction representation 
for the one-fermion diagonal Green functions immediately reproduces 
the results for the spin-$\frac{1}{2}$ transverse isotropic $XY$ chains 
having periods 2 and 3 and in contrast to other approaches easily yields the 
results for larger periods (e.g., 4 and 12). 
It should be stressed that the elaborated approach is a systematic method 
that permits to consider in the same fashion the regularly nonuniform chains
with randomness that cannot be done within the frames of the approaches 
exploited in Refs. 12-27. We present a comprehensive study of the 
thermodynamic properties (density of states, gap in the energy spectrum,
entropy, specific heat, magnetization, susceptibility) of regularly 
alternating chains having periods 2, 3, 4, and 12
discussing in detail the dependences of the energy gap and
the low-temperature transverse
magnetization on the transverse field and comparing the latter dependence 
with the corresponding
one for the classical spin chain. We underline a 
possibility of a nonzero transverse magnetization at the zero average 
transverse field in the periodic nonuniform chain 
owing to a regular
nonuniformity. This material constitutes Section 2. In Section 3 we 
demonstrate the influence of a diagonal disorder on the effects caused by 
periodic 
nonuniformity assuming the transverse fields to be independent random  
Lorentzian variables. For simplicity we restrict ourselves by the case
of a 
chain having period 2. The results obtained for the spin-$\frac{1}{2}$ 
transverse isotropic $XY$
chains with period 2 are applied for an analysis
of the spin-Peierls instability
in the adiabatic limit
with respect to dimerization 
in the presence of a non-random/random (Lorentzian) transverse field 
(Section 4).
We find how
the (random) transverse field
influences the dependence
of the (averaged) total energy
on the dimerization parameter
tracing a suppression of dimerization by
the non-random (random) field.

\section{Periodic nonuniform spin-$\frac{1}{2}$
	isotropic $XY$ chain in a transverse field}

Let us consider a cyclic
nonuniform isotropic $XY$ chain of $N$
(eventually $N\to\infty$) spins $s=\frac{1}{2}$
in a transverse field. The Hamiltonian of the system reads
\begin{eqnarray}
H=\sum_{n=1}^{N}\Omega_ns_n^z
+2\sum_{n=1}^{N}I_n\left(s^x_ns^x_{n+1}+s^y_ns^y_{n+1}\right)
\nonumber\\
=\sum_{n=1}^{N}\Omega_n\left(s_n^+s_n^--\frac{1}{2}\right)
+\sum_{n=1}^{N}I_n\left(s^+_ns^-_{n+1}+s^-_ns^+_{n+1}\right),
\;\;\;
s^{\alpha}_{n+N}=s^{\alpha}_{n}.
\end{eqnarray}
Here $\Omega_n$ is the transverse field at site $n$
and $2I_n$ is the
exchange coupling between the sites $n$ and $n+1$.
Let us note that 
$s^z=\sum_{n=1}^Ns_n^z$
commutes with the Hamiltonian $H$ (1)
and hence
the eigenfunctions of $H$ can be classified according to 
eigenvalues of 
$s^z$.
Moreover, at $\Omega_n=0$
the ground state of $H$ corresponds to $s^z=0$.
After making use of the Jordan-Wigner transformation
one comes to a cyclic chain of spinless fermions
governed by the Hamiltonian
\begin{eqnarray}
H=\sum_{n=1}^{N}\Omega_n\left(c_n^+c_n-\frac{1}{2}\right)
+\sum_{n=1}^{N}I_n\left(c^+_nc_{n+1}-c_nc^+_{n+1}\right).
\end{eqnarray}
The so-called boundary term is not
essential for calculation of thermodynamic functions$^{30}$
and has been omitted.
We shall discuss the most general case, i.e. 
assuming that
both transverse fields
and exchange couplings vary from site to site.
Note, that 
in the particular case when
the transverse field is uniform
one recognizes in Eq. (2) the Hamiltonian of the system
considered in Ref. 2.
In addition, in another limiting case after substitution
$\Omega_n\rightarrow Uw_n,$
$I_n\rightarrow -t$
Eq. (2) transforms into the
Hamiltonian of a one-dimensional spinless Falicov-Kimball model
in the notations used in Refs. 4, 5.

Let us introduce the temperature double-time Green functions
\linebreak
$G_{nm}^{\mp}(t)
=\mp{\mbox{i}}\theta(\pm t)
\langle\left\{c_n(t), c_m^+(0)\right\}\rangle,$
$G_{nm}^{\mp}(t)=\left(1/2\pi\right)
\int_{-\infty}^{\infty}{\mbox{d}}\omega
\exp\left(-{\mbox{i}}\omega t\right)
G_{nm}^{\mp}(\omega\pm{\mbox{i}}\epsilon),$
$\epsilon\rightarrow+0,$
where the angular brackets denote the thermodynamic average.
Consider further the set of equations
of motion
for
$G_{nm}^{\mp}\equiv G_{nm}^{\mp}(\omega\pm{\mbox{i}}\epsilon)$
\begin{eqnarray}
\left(\omega\pm{\mbox{i}}\epsilon-\Omega_n\right)G_{nm}^{\mp}
-I_{n-1}G_{n-1,m}^{\mp}
-I_{n}G_{n+1,m}^{\mp}
=\delta_{nm}.
\end{eqnarray}
Our task is to evaluate the diagonal Green functions
$G_{nn}^{\mp}$,
the imaginary part of which gives the density of states
$\rho(\omega)$,
\begin{eqnarray}
\rho(\omega)
=\mp\frac{1}{\pi N}
\sum_{n=1}^N{\mbox{Im}}G_{nn}^{\mp},
\end{eqnarray}
that on its part, yields the thermodynamic properties of
the spin model (1).

It is a simple matter to obtain
from Eq. (3) the following representation for $G_{nn}^{\mp}$
\begin{eqnarray}
G_{nn}^{\mp}
=\frac{1}
{\omega\pm{\mbox{i}}\epsilon-\Omega_n-\Delta^-_n-\Delta^+_n},
\nonumber\\
\Delta^-_n=\frac{I_{n-1}^2}
{\omega\pm{\mbox{i}}\epsilon-\Omega_{n-1}-
\frac{I_{n-2}^2}
{\omega\pm{\mbox{i}}\epsilon-\Omega_{n-2}-_{\ddots}}},
\nonumber\\
\Delta^+_n=\frac{I_{n}^2}
{\omega\pm{\mbox{i}}\epsilon-\Omega_{n+1}-
\frac{I_{n+1}^2}
{\omega\pm{\mbox{i}}\epsilon-\Omega_{n+2}-_{\ddots}}}.
\end{eqnarray}
Equations (4), (5) are extremely useful
for examining thermodynamic properties
of the {\em periodic} nonuniform
spin-$\frac{1}{2}$ isotropic $XY$ chain in a transverse field,
since the evaluation of periodic continued fractions$^{31}$ emerging in 
(5)
is quite simple
and reduces to solving quadratic equations.

It should be noted here that the continued-fraction representation
of the one-particle Green functions has been widely used for
tight-binding electrons over the last two decades.
As an example let us refer
here to the papers of Haydock, Heine and Kelly$^{32,33}$
and the review articles.$^{34,35}$
However, those studies were aimed mainly on
getting the electronic band structure of non-translationally invariant
systems (alternatively to the band theory)
starting from the local environment of atom
and in practice were connected with an appropriate approximative
termination
of continued fractions.
In what follows we shall use the exact values of continued fractions
(since they are periodic)
to reveal the effects of regular nonuniformity on
the magnon band structure.

Consider at first a uniform chain
$\Omega_0I\Omega_0I\ldots\;$.
In this case one comes to
a periodic continued fraction having a period 1
\begin{eqnarray}
\Delta^-_n
=\Delta^+_n
=\Delta
=\frac{I^2}
{\omega\pm{\mbox{i}}\epsilon-\Omega_0-
\frac{I^2}
{\omega\pm{\mbox{i}}\epsilon-\Omega_0-_{\ddots}}}
=\frac{I^2}
{\omega\pm{\mbox{i}}\epsilon-\Omega_0 -\Delta}.
\end{eqnarray}
The quadratic equation for $\Delta$ (6) can be solved with
\begin{eqnarray}
\Delta
=\left\{\frac{1}{2}
\left[
\omega\pm{\mbox{i}}\epsilon-\Omega_0
+\sqrt{\left(\omega\pm{\mbox{i}}\epsilon-\Omega_0\right)^2-4I^2}
\right],\;
\frac{1}{2}
\left[
\omega\pm{\mbox{i}}\epsilon-\Omega_0
-\sqrt{\left(\omega\pm{\mbox{i}}\epsilon-\Omega_0\right)^2-4I^2}
\right]\right\}
\end{eqnarray}
and therefore
$\rho(\omega)$
according to (4), (5) becomes
\begin{eqnarray}
\rho(\omega)
=\left\{
\begin{array}{ll}
\frac{1}{\pi}
\frac{1}{\sqrt{4I^2-\left(\omega-\Omega_0\right)^2}},
& {\mbox{if}}\;\;\;4I^2-(\omega-\Omega_0)^2>0,
\\
0,
& {\mbox{otherwise.}}
\end{array}
\right.
\end{eqnarray}
The self-consistent equation for the continued fraction (6)
introduces a spurious root.
However, the false solution is eliminated
requiring $\rho(\omega)$ to be not negative.
Let us emphasize the attractive features of
the continued-fraction approach
reminding how
$\rho(\omega)$ (8)
can be obtained within the frames
of the standard technique.
Usually one substitutes into Eq. (3)
$G_{nm}^{\mp}=(1/N)\sum_{\kappa}
\exp[{\mbox{i}}(n-m)\kappa]G_{\kappa}^{\mp}$
to obtain
$G_{\kappa}^{\mp}
=1/(\omega\pm{\mbox{i}}\epsilon-\Omega_0-2I\cos\kappa)$
and then evaluates the integral
$G_{nn}^{\mp}
=(1/2\pi)\int_{-\pi}^{\pi}{\mbox{d}}\kappa
/(\omega\pm{\mbox{i}}\epsilon-\Omega_0-2I\cos\kappa)$
using, for example, contour integrals
to get
$G_{nn}^{\mp}=1
/\sqrt{\left(\omega\pm{\mbox{i}}\epsilon-\Omega_0\right)^2-4I^2}$
and therefore
the density of states (8).

The advantages of the continued-fraction approach
become clear while treating the periodic nonuniform chains.
We shall demonstrate this in some 
detail for regularly modulated chains with
periods of modulation  of 2, 3 and 4.

(i)
Consider a regular alternating chain
$\Omega_1I_1\Omega_2I_2\Omega_1I_1\Omega_2I_2\ldots\;$.
In this case
periodic continued fractions having a period 2 emerge.
Solving similar quadratic equations as (6)
for
$\Delta_n^-,$
$\Delta_n^+,$
$\Delta_{n+1}^-,$
$\Delta_{n+1}^+$
one obtains as a result the Green functions
$G_{nn}^{\mp},$
$G_{n+1,n+1}^{\mp}$
and therefore the density of states
$\rho(\omega)$
\begin{eqnarray}
\rho(\omega)
=\left\{
\begin{array}{ll}
\frac{1}{2\pi}
\frac{\mid 2\omega-\Omega_1-\Omega_2\mid}
{\sqrt{{\cal{B}}(\omega)}},
& {\mbox{if}}\;\;\;{\cal{B}}(\omega)>0,
\\
0,
& {\mbox{otherwise;}}
\end{array}
\right.
\nonumber\\
{\cal{B}}(\omega)
=4I_1^2I_2^2
-\left[
\left(\omega-\Omega_1\right)
\left(\omega-\Omega_2\right)
-I_1^2-I_2^2
\right]^2
\nonumber\\
=-\left(\omega-b_1\right)
\left(\omega-b_2\right)
\left(\omega-b_3\right)
\left(\omega-b_4\right).
\end{eqnarray}
Here
$b_1\ge b_2\ge b_3\ge b_4$
denote the four roots of the equation
${\cal{B}}(\omega)=0$, namely
\begin{eqnarray}
\left\{b_i\right\}
=
\left\{
\frac{1}{2}
\left(
\Omega_1+\Omega_2
\right)
\pm{\sf{b}}_1,
\;\;\;
\frac{1}{2}
\left(
\Omega_1+\Omega_2
\right)
\pm{\sf{b}}_2
\right\}
\end{eqnarray}
with
${\sf{b}}_1=\frac{1}{2}\sqrt{\left(\Omega_1-\Omega_2\right)^2
+4\left(\vert I_1\vert+\vert I_2\vert\right)^2}$,
${\sf{b}}_2=\frac{1}{2}\sqrt{\left(\Omega_1-\Omega_2\right)^2
+4\left(\vert I_1\vert-\vert I_2\vert\right)^2}$.
Solving the inequality
${\cal{B}}(\omega)>0$
one can write the density of states
$\rho(\omega)$ (9) in the explicit form
\begin{eqnarray}
\rho(\omega)
=\left\{
\begin{array}{ll}
0,
& {\mbox{if}}\;\;\;\omega<b_4,\;\;\;b_3<\omega<b_2,\;\;\;b_1<\omega,
\\
\frac{1}{2\pi}
\frac{\mid 2\omega-\Omega_1-\Omega_2\mid}
{\sqrt{{\cal{B}}(\omega)}},
& {\mbox{if}}\;\;\;b_4<\omega<b_3,\;\;\;b_2<\omega<b_1.
\end{array}
\right.
\end{eqnarray}
The result for
the uniform chain (8)
is contained
in the density of states (11), (10), (9)
as a partial case
when $\Omega_1=\Omega_2=\Omega_0$,
$I_1=I_2=I$.

(ii) Next we consider the regularly modulated   chain
$\Omega_1I_1\Omega_2I_2\Omega_3I_3\Omega_1I_1\Omega_2I_2\Omega_3I_3\ldots\;$.
In this case 
one generates
the corresponding periodic continued fractions of period 3.
Going along the lines as described above one gets
\begin{eqnarray}
\rho(\omega)
=\left\{
\begin{array}{ll}
\frac{1}{3\pi}
\frac{\mid
I_1^2+I_2^2+I_3^2
-\left(\omega-\Omega_1\right)\left(\omega-\Omega_2\right)
-\left(\omega-\Omega_1\right)\left(\omega-\Omega_3\right)
-\left(\omega-\Omega_2\right)\left(\omega-\Omega_3\right)
\mid}
{\sqrt{{\cal{C}}(\omega)}},
& {\mbox{if}}\;\;\;{\cal{C}}(\omega)>0,
\\
0,
& {\mbox{otherwise;}}
\end{array}
\right.
\nonumber\\
{\cal{C}}(\omega)
=4I_1^2I_2^2I_3^2
-\left[
I_1^2\left(\omega-\Omega_3\right)
+I_2^2\left(\omega-\Omega_1\right)
+I_3^2\left(\omega-\Omega_2\right)
-\left(\omega-\Omega_1\right)
\left(\omega-\Omega_2\right)
\left(\omega-\Omega_3\right)
\right]^2
\nonumber\\
=-\prod_{j=1}^6\left(\omega-c_j\right),
\end{eqnarray}
where $c_j$ are the six roots of the equation ${\cal{C}}(\omega)=0$.
To find them one must solve two qubic equations
that follow from Eq. (12).

(iii) Finally, let us
consider the regularly modulated chain
$\Omega_1I_1\Omega_2I_2\Omega_3I_3\Omega_4I_4
\Omega_1I_1\Omega_2I_2\Omega_3I_3\Omega_4I_4\ldots\;$.
In this case one gets
periodic continued fractions with period 4.
The density of states for such a chain is given by
\begin{eqnarray}
\rho(\omega)
=\left\{
\begin{array}{ll}
\frac{1}{4\pi}
\frac{\mid{\cal{W}}(\omega)\mid}
{\sqrt{{\cal{D}}(\omega)}},
& {\mbox{if}}\;\;\;{\cal{D}}(\omega)>0,
\\
0,
& {\mbox{otherwise;}}
\end{array}
\right.
\nonumber\\
{\cal{W}}(\omega)
=I_1^2(2\omega-\Omega_3-\Omega_4)
+I_2^2(2\omega-\Omega_1-\Omega_4)
+I_3^2(2\omega-\Omega_1-\Omega_2)
+I_4^2(2\omega-\Omega_2-\Omega_3)
\nonumber\\
-(\omega-\Omega_1)(\omega-\Omega_2)(\omega-\Omega_3)
-(\omega-\Omega_1)(\omega-\Omega_2)(\omega-\Omega_4)
\nonumber\\
-(\omega-\Omega_1)(\omega-\Omega_3)(\omega-\Omega_4)
-(\omega-\Omega_2)(\omega-\Omega_3)(\omega-\Omega_4),
\nonumber\\
{\cal{D}}(\omega)
=4I_1^2I_2^2I_3^2I_4^2
-\left[
\left(\omega-\Omega_1\right)
\left(\omega-\Omega_2\right)
\left(\omega-\Omega_3\right)
\left(\omega-\Omega_4\right)
\right.
\nonumber\\
\left.
-I_1^2\left(\omega-\Omega_3\right)\left(\omega-\Omega_4\right)
-I_2^2\left(\omega-\Omega_1\right)\left(\omega-\Omega_4\right)
\right.
\nonumber\\
\left.
-I_3^2\left(\omega-\Omega_1\right)\left(\omega-\Omega_2\right)
-I_4^2\left(\omega-\Omega_2\right)\left(\omega-\Omega_3\right)
\right.
\nonumber\\
\left.
+I_1^2I_3^2+I_2^2I_4^2
\right]^2
=-\prod_{j=1}^8\left(\omega-d_j\right),
\end{eqnarray}
where $d_j$ are the eight roots of the equation
${\cal{D}}(\omega)=0$.
To find them one must solve two equations of 4th order
that follow from Eq. (13).
Let us note that all
$d_j$
(as well as all $c_j$)
are real
since they can be viewed as eigenvalues of symmetric matrices.$^2$

There are no principal difficulties in proceeding the
analytic
calculations of $\rho(\omega)$
for larger periods,
except the fact that they become more cumbersome.
All
Green functions
required for getting the density of states $\rho(\omega)$ (4)
are calculated by solving quadratic equations,
however, further analysis of the band structure is becoming
more complicated.
This analysis, however, can be easily implemented on a computer
and the results for the chains having period 12 presented below
were obtained in such a manner.

Let us discuss the
results for the density of states
for the considered periodic
nonuniform chains.
The main consequence of introducing the nonuniformity
is a splitting of the initial magnon band
into several subbands
(compare (8) and (9) - (13)).
The edges of the subbands are determined by the roots of equations
${\cal{B}}(\omega)=0$,
${\cal{C}}(\omega)=0$,
${\cal{D}}(\omega)=0$, etc..
$\rho(\omega)$ is positive inside the subbands,
tends to infinity inversely proportionally to the square root of
$\omega-\omega_e$
when $\omega$ approaches the subbands edges $\omega_e$,
and is equal to zero outside the subbands.
The number of subbands does not exceed the period of the chain.
At special (symmetric) values of the Hamiltonian parameters the roots of
the equation
that determines the subband edges may become multiple
and the zeros in the denominator and the numerator in the expression for
$\rho(\omega)$
may cancel each other.
As a result due to an increase of symmetry one may observe a smaller
number of subbands.
This `mechanism' is easily traced, for example,
in formulas (9) - (11)
if putting $\Omega_1=\Omega_2$, $\vert I_1\vert=\vert I_2\vert$.
The described magnon band structure
can be seen in
Figs. 1 and 2a, 3 and 4a, and 5 and 6a
where we show
$\rho(\omega)$ for
a few particular periodic nonuniform chains
having periods 2, 3, and 12, respectively.
The splitting caused by periodic nonuniformity
in fact
is not surprising.
The periodic nonuniform chain
is simply another viewpoint on
the uniform chain with a crystalline unit cell
containing several sites.
On the other hand,
it is generally known that
one may expect several subbands for a crystal having several
atoms per unit cell.$^{36}$

Further
one can easily calculate the widths of energy gaps
in the magnon spectrum
that appear due to nonuniformity.
For example, for a chain having a period 2 one finds
\begin{eqnarray}
b_2-b_3
=\sqrt{(\Omega_1-\Omega_2)^2+4(\vert I_1\vert-\vert I_2\vert)^2}.
\end{eqnarray}
This quantity is connected with a gap
between the ground state energy and the first excited state energy
of the spin chain.
As an example consider the chain 
$\Omega_0I_1\Omega_0I_2
\Omega_0I_1\Omega_0I_2\ldots\;$.
The edges of the upper magnon subband are given by
$\Omega_0+\vert I_1\vert+\vert I_2\vert$
and
$\Omega_0+\vert\vert I_1\vert-\vert I_2\vert\vert$,
whereas the edges of the lower magnon subband are given by
$\Omega_0-\vert\vert I_1\vert-\vert I_2\vert\vert$,
and
$\Omega_0-\vert I_1\vert-\vert I_2\vert$.
At $\Omega_0=0$ the ground state (for which $s^z=0$)
corresponds to the filled lower subband 
and the empty upper subband 
and the energy spectrum exhibits a gap 
$\Delta(0)=\vert\vert I_1\vert-\vert I_2\vert\vert$
(this is the energy required to create a hole in the lower subband ---
the first excited state of the spin chain
(with $s^z\ne 0$)).
With increasing of $\Omega_0$ 
the gap
$\Delta$ decreases as
$\Delta(\Omega_0)=\vert\vert I_1\vert-\vert I_2\vert\vert-\Omega_0$
and becomes zero at
$\Omega_0=\vert\vert I_1\vert-\vert I_2\vert\vert$.
With further increasing of $\Omega_0$ the gap remains equal to zero up to 
the value of the transverse field  
$\Omega_0=\vert I_1\vert+\vert I_2\vert$
after which the gap opens and increases as
$\Delta(\Omega_0)=\Omega_0-\vert I_1\vert-\vert I_2\vert$
(the ground state for the transverse field larger than 
$\vert I_1\vert+\vert I_2\vert$
corresponds to the empty subbands and
the written
$\Delta(\Omega_0)$ is the energy
required to create a particle in the vicinity of the lower edge of the 
lower subband). For chains with larger periods one finds more complicated 
behaviour of the energy gap $\Delta$ with varying of the field $\Omega_0$
(see Fig. 7a and Figs. 7b - 7d).

The splitting of the magnon band into subbands caused by nonuniformity
has interesting consequences for thermodynamic properties.
The entropy, specific heat, transverse magnetization and
static transverse linear susceptibility are determined through the density
of states
according to the following formulas
\begin{eqnarray}
s=\int\limits_{-\infty}^{\infty}
{\mbox{d}}E\rho(E)
\left[\ln\left(2\cosh\frac{E}{2kT}\right)-
\frac{E}{2kT}\tanh \frac{E}{2kT}\right],
\\
c=\int\limits_{-\infty}^{\infty}
{\mbox{d}}E\rho(E)
\left(
\frac{\frac{E}{2kT}}
{\cosh\frac{E}{2kT}}
\right)^2,
\\
m_z=-\frac{1}{2}\int\limits_{-\infty}^{\infty}
{\mbox{d}}E\rho(E)\tanh\frac{E}{2kT},
\\
\chi_{zz}=-\frac{1}{kT}\int\limits_{-\infty}^{\infty}
{\mbox{d}}E\rho(E)\frac{1}{4\cosh^2\frac{E}{2kT}}.
\end{eqnarray}

Apparently the most spectacular changes caused by regular nonuniformity
are observed in the dependence of transverse magnetization (17)
on transverse field at low temperatures
(Figs. 2b, 4b, 6b).
Since for $T\rightarrow 0$,
$\tanh\frac{E}{2kT}$ tends either to $-1$ if $E<0$,
or to $1$ if $E>0$,
one immediately finds
due to the splitting of the magnon band into 
subbands
that the low-temperature dependence of
$m_z$ versus $\Omega_0$
must be composed of sharply increasing parts
(they appear when $E=0$ moves with increasing of $\Omega_0$
from the  bottom to the top of each subband)
separated by horizontal parts
(they appear when $E=0$ moves with increasing of $\Omega_0$
inside the gaps).
The number of plateaus is determined by the number of subbands.
It should be emphasized here that a study of magnetization plateaus for 
quantum spin chains is a hot topic at the present time.$^{37}$
However in such studies usually more general spin chains are attacked
which cannot be treated within the frames of the described approach. For 
example, the spin-$\frac{1}{2}$ $XXX$ chain can be mapped onto the chain of 
{\it {interacting}} 
spinless fermions with the intersite interaction of 
the same order as the hopping integral and hence the results derived 
rigorously for noninteracting fermions cannot be immediately extended for 
this more complicated spin chain.

It is interesting to note that the appearance of plateaus in the dependence
of transverse magnetization on transverse field at $T=0$
for the regularly nonuniform isotropic $XY$ chains
essentially differs in the quantum and classical
cases.
The Hamiltonian of the classical nonuniform
isotropic $XY$ chain in a transverse field reads
\begin{eqnarray}
H=\sum_{n=1}^N\Omega_ns\cos\theta_n
+2\sum_{n=1}^NI_ns^2\cos(\phi_n-\phi_{n+1})
\sin\theta_n
\sin\theta_{n+1}
\end{eqnarray}
that
immediately yields the ansatz for the ground state energy
in the uniform case
\begin{eqnarray}
E_0
=N\Omega_0 s\cos\theta-2 N\vert I\vert s^2\sin^2\theta
=N\Omega_0 m_z+2 N\vert I\vert\left(m_z^2-s^2\right)
\end{eqnarray}
where the ground state transverse magnetization
$m_z=s\cos\theta$ has been introduced.
Minimizing $E_0$ with respect to
$\cos\theta$ one finds
that
for $s=\frac{1}{2}$
the quantity
$-m_z$ increases as $\frac{1}{2}\frac{\Omega_0}{2\vert I\vert}$
while $\Omega_0$ increases from $0$ to
$2\vert I\vert$ and
$-m_z=\frac{1}{2}$
with further increase of $\Omega_0$.
Using numerical calculations for finite chains
(the number of spins $N$ is a multiple of 12)
with periods 2, 3, 12
we found that
the detailed profiles for the quantum and classical chains
are different,
although the values of the transverse field
at which a saturation of the transverse
magnetization occurs are the same.
Though one could argue that the magnetization plateaus are connected with
the quantum nature of the spins we found
for special parameter sets even in the classical chain
plateaus
in the dependence $-m_z$ versus $\Omega_0$
(compare
dashed curves
in Figs. 8a, 8b and
in Figs. 2b, 4b).
For instance,
the well pronounced plateau shown by the dashed line in Fig. 8b occurs
at the same height as in the quantum case. 
The corresponding classical state
is a state 
$\downarrow \uparrow \uparrow \downarrow \uparrow \uparrow 
\downarrow \uparrow \uparrow  \downarrow \uparrow \uparrow \ldots $ 
where
the arrows symbolize classical spins pointing either 
in $-z$- or $+z$-direction.
An
evident difference between the quantum and classical case is connected with
the slope of the $m_z(\Omega_0)$ curve at $T=0$. 
The slope remains finite in the
classical case but becomes infinite approaching the plateaus in the quantum
case. The infinite slope in the quantum case is clearly a
consequence of the singularities in the density of states.

One of the interesting magnetic properties of the periodic nonuniform
spin-$\frac{1}{2}$ isotropic $XY$ chain is the possibility
of the existence of a non-zero transverse magnetization
$m_z$ at zero average transverse field
($\sum_{n=1}^N\Omega_n=0$). For illustration we consider as
an example a chain having
the period 4 and the parameters $\Omega_1=\Omega_3=0$,
$\Omega_2=-\Omega_4<0$,
$\vert I_1\vert =\vert I_2\vert >0$,
$\vert I_3\vert =\vert I_4\vert =0$.
At site $n+1$ we have the transverse field
$\Omega_2<0$ surrounded on the left and right side by the strong
couplings 
$\vert I_1\vert =\vert I_2\vert$. 
At site $n+3$ we have the
transverse field $-\Omega_2 >0$ surrounded by the weak couplings
$\vert I_3\vert =\vert I_4\vert =0$.
One may expect that the local transverse
magnetization at site $n+1$ has
a smaller value and opposite direction
with respect to that quantity at site $n+3$
and therefore a non-zero total transverse magnetization
at zero average transverse field may be expected.
Consider the described chain in more detail.
From Eq. (13) for the above set of parameters
it follows that
\begin{eqnarray}
\rho(\omega)
=\lambda_1\delta\left(\omega-
\frac{\Omega_2-\sqrt{\Omega_2^2+8I_1^2}}{2}\right)
+\lambda_2\delta\left(\omega\right)
+\lambda_3\delta\left(\omega-
\frac{\Omega_2+\sqrt{\Omega_2^2+8I_1^2}}{2}\right)
+\lambda_4\delta\left(\omega+\Omega_2\right)
\end{eqnarray}
and the coefficients $\lambda_j$ may be found comparing (21) and (13)
in the vicinity of
$\frac{\Omega_2-\sqrt{\Omega_2^2+8I_1^2}}{2}$,
$0$,
$\frac{\Omega_2+\sqrt{\Omega_2^2+8I_1^2}}{2}$
and
$-\Omega_2$.
As a result one gets $\lambda_j=\frac{1}{4}$
(see Fig. 9a).
Now transverse magnetization (17) at $T=0$ is
$m_z=-\frac{1}{8}\ne 0$
although $\sum_{n=1}^N\Omega_n=0$
(solid curves in Figs. 9b, 9c;
in the latter figure the solid curve especially clearly shows that
$-m_z=\frac{1}{8}$).
If $\vert I_3\vert=\vert I_4\vert\ne 0$ 
the magnon subbands look as in Fig. 9a
and at $T=0$ one has $m_z=0$ (Figs. 9b, 9c).
However, such a position of the subbands
provides an interesting temperature dependence
of $m_z$
at $\sum_{n=1}^N\Omega_n=0$
(dashed and dotted curves in Fig. 9c)
reminding the `order from disorder' phenomenon,$^{38-40}$
i.e. increasing of order with increasing temperature.

Let us turn to other thermodynamic quantities.
Every infinite slope in the dependence
$m_z$ versus $\Omega_0$ at $T=0$
induces a singularity in the dependence
$\chi_{zz}$ versus $\Omega_0$ at $T=0$.
However, there is no need to plot this dependence.
Since $1/4kT\cosh^2\frac{E}{2kT}$
tends to $\delta(E)$ as $T\to0$
one gets from (18) that at $T=0$
$\;$
$-\chi_{zz}=\rho(0)$.
The latter dependence as a matter of fact can be seen
in Figs. 2a, 4a, 6a.
The changes
in the temperature dependences of entropy and specific heat
due to nonuniformity
which are
displayed in Figs. 2c, 4c, 6c
and 2d, 4d, 6d
can be understood while bearing in mind
the behaviour of integrands in (15), (16)
that are products of
the functions with evident dependences on the temperature
and the density of states.
Note that as a result of the magnon band splitting
the temperature dependence of the specific heat may
exhibit a two-peak structure (Fig. 2d)
or even a more complicated behaviour (solid curve in Fig. 4d).
Finally we look at $\chi_{zz}$. As mentioned above
at $T=0$ we have
$\;$
$-\chi_{zz}=\rho(0)$.
Analysing the density of states depicted in Figs. 2a, 4a, 6a
one finds that nonuniformity may either suppress or enhance
the initial (that is at $\Omega_0=0$)
static transverse linear susceptibility
$-\chi_{zz}$ at $T=0$
shown in
Figs. 2f, 4f, 6f.

\section{Periodic nonuniform spin-$\frac{1}{2}$
isotropic $XY$ chain
in a random Lo\-rentzian transverse field}

In this Section we consider a generalization of model (1) including 
additional
randomness in the transverse fields. We assume
the transverse fields to be independent random variables
each with a Lorentzian probability distribution
\begin{eqnarray}
p(\Omega_n)=\frac{1}{\pi}
\frac{\Gamma_n}
{\left(\Omega_{0n}-\Omega_n\right)^2+\Gamma_n^2}.
\end{eqnarray}
Here $\Omega_{0n}$  is the mean value of the transverse field at site
$n$ and $\Gamma_n$ is the width of its distribution.
We are interested in the random-averaged density of states
$\overline{\rho(\omega)}$
that follows from the random-averaged diagonal Green functions
$\overline{G_{nn}^{\mp}}$ according to Eq. (4).
Repeating the arguments presented in Refs. 6-11
one gets the following set of equations for the random-averaged
Green functions
\begin{eqnarray}
\left(\omega\pm{\mbox{i}}\Gamma_n
-\Omega_{0n}\right)
\overline{G_{nm}^{\mp}}
-I_{n-1}\overline{G_{n-1,m}^{\mp}}
-I_{n}\overline{G_{n+1,m}^{\mp}}
=\delta_{nm}
\end{eqnarray}
that immediately yields
\begin{eqnarray}
\overline{G_{nn}^{\mp}}
=\frac{1}
{\omega\pm{\mbox{i}}\Gamma_n
-\Omega_{0n}-\Delta^-_n-\Delta^+_n},
\nonumber\\
\Delta^-_n=\frac{I_{n-1}^2}
{\omega\pm{\mbox{i}}\Gamma_{n-1}
-\Omega_{0,n-1}-
\frac{I_{n-2}^2}
{\omega\pm{\mbox{i}}\Gamma_{n-2}
-\Omega_{0,n-2}-_{\ddots}}},
\nonumber\\
\Delta^+_n=\frac{I_{n}^2}
{\omega\pm{\mbox{i}}\Gamma_{n+1}
-\Omega_{0,n+1}-
\frac{I_{n+1}^2}
{\omega\pm{\mbox{i}}\Gamma_{n+2}
-\Omega_{0,n+2}-_{\ddots}}}.
\end{eqnarray}
In case
$\Omega_{0n},$
$\Gamma_n,$
$I_n$
vary regularly from site to site
one again comes to the periodic
continued fractions.
They can be calculated as solutions of the corresponding
quadratic equations.
Thus one gets rigorously the random-averaged Green functions
and therefore the random-averaged density of states.
For example, for a regular random chain
$\Omega_{01}\Gamma_1I_1\Omega_{02}\Gamma_2I_2
\Omega_{01}\Gamma_1I_1\Omega_{02}\Gamma_2I_2
\ldots$
one finds
\begin{eqnarray}
\overline{\rho(\omega)}
=\frac{1}{2\pi}\frac{\vert{\cal{Y}}(\omega)\vert}
{{\cal{B}}(\omega)};
\nonumber\\
{\cal{Y}}(\omega)
=(\Gamma_1+\Gamma_2)
\sqrt{\frac{{\cal{B}}(\omega)+{\cal{B}}^{\prime}(\omega)}{2}}
-{\mbox{sgn}}{\cal{B}}^{\prime\prime}(\omega)
(2\omega-\Omega_{01}-\Omega_{02})
\sqrt{\frac{{\cal{B}}(\omega)-{\cal{B}}^{\prime}(\omega)}{2}},
\nonumber\\
{\cal{B}}(\omega)
=\sqrt{\left({\cal{B}}^{\prime}(\omega)\right)^2
+\left({\cal{B}}^{\prime\prime}(\omega)\right)^2},
\nonumber\\
{\cal{B}}^{\prime}(\omega)
=\left[
(\omega-\Omega_{01})
(\omega-\Omega_{02})
-\Gamma_1\Gamma_2-I_1^2-I_2^2
\right]^2
-
\left[
(\omega-\Omega_{01})\Gamma_2
+
(\omega-\Omega_{02})\Gamma_1
\right]^2
-4I_1^2I_2^2,
\nonumber\\
{\cal{B}}^{\prime\prime}(\omega)
=2\left[
(\omega-\Omega_{01})
(\omega-\Omega_{02})
-\Gamma_1\Gamma_2-I_1^2-I_2^2
\right]
\left[
(\omega-\Omega_{01})\Gamma_2
+
(\omega-\Omega_{02})\Gamma_1
\right].
\end{eqnarray}
The random-averaged density of states (25)
transforms into (9)
if $\Gamma_1=\Gamma_2=0$,
and into the result reported in Ref. 9,
$\overline{\rho(\omega)}
=\mp(1/\pi){\mbox{Im}}
1/\sqrt{(\omega\pm{\mbox{i}}\Gamma-\Omega_0)^2-4I^2},$
if
$\Omega_{01}=\Omega_{02}=\Omega_{0},$
$\Gamma_1=\Gamma_2=\Gamma,$
$I_1=I_2=I.$

Let us discuss the effects of
the considered diagonal Lorentzian disorder.
The main effect of the randomness
is smearing out the band structure.
However, one can see a difference
in smoothed magnon subbands for the uniform disorder
(when $\Gamma_1=\Gamma_2$) (see Fig. 10a)
and the nonuniform disorder
(when $\Gamma_1\ne\Gamma_2$) (see Fig. 11a).
Namely, in the former case both subbands are smeared out in the same way,
whereas in the latter case,
the subbands are smeared out differently and,
at least for small strengths of disorder, in one subband the peaks at
the band edges persist.
This circumstance in the latter case induces an interesting 
step-like behaviour of the low-temperature
transverse magnetization as a function of transverse field.
Namely, as can be seen in Fig. 11b
the disorder smooths only one step
in contrast to Fig. 10b in which both steps are smeared out.
The difference in the influence of the uniform and nonuniform
disorders on other thermodynamic quantities
can be seen in Figs. 10c - 10f and 11c - 11f.

%% file: lrb07.tex
\section{Periodic nonuniform spin-$\frac{1}{2}$ isotropic $XY$ chains
	and spin-Peierls instability}

In this Section we want to demonstrate that the results
for the density of states of the 
periodic nonuniform 
spin-$\frac{1}{2}$ isotropic $XY$ chains
obtained
within
the continued-fraction approach may be of use
for the  study of the spin-Peierls instability in these chains
in adiabatic limit.
The discovery of existence of the spin-Peierls transition in
the inorganic compound CuGeO$_3$$^{41,42}$
has stimulated much research work in this field.
In particular, the influence of an external field
or randomness attracts much interest both from experimental and
theoretical viewpoints (see e.g. Refs. 42-49).

Let us start from the non-random case.
In order to examine 
the instability 
of the spin chain 
with respect to dimerization
one must calculate the ground state energy per site
of the regularly alternating chain
$\Omega_1I_1\Omega_2I_2\Omega_1I_1\Omega_2I_2\ldots$
(see Eqs. (9) - (11))
\begin{eqnarray}
e_0=
-\frac{1}{2}\int\limits_{-\infty}^{\infty}
{\mbox{d}}E\rho(E)\vert E\vert
=-\frac{1}{2\pi}
\int\limits_{-{\sf{b}}_1}^{-{\sf{b}}_2}
{\mbox{d}}E\frac{\vert E\vert
\left(
\vert E-\hat{\Omega}\vert+\vert E+\hat{\Omega}\vert
\right)}
{\sqrt{-\left(E^2-{\sf{b}}_1^2\right)
\left(E^2-{\sf{b}}_2^2\right)}}
\end{eqnarray}
where $\hat{\Omega}=(\Omega_1+\Omega_2)/2$.
Depending on the value of $\hat{\Omega}$
formula (26) can be rewritten as follows
\begin{eqnarray}
e_0=
-\frac{1}{\pi}
\int\limits_{-{\sf{b}}_1}^{-{\sf{b}}_2}
{\mbox{d}}E\frac{\vert\hat{\Omega}\vert \vert E\vert}
{\sqrt{-\left(E^2-{\sf{b}}_1^2\right)
\left(E^2-{\sf{b}}_2^2\right)}}
\end{eqnarray}
if ${\sf{b}}_1\le \vert\hat{\Omega}\vert$,
\begin{eqnarray}
e_0=
-\frac{1}{\pi}
\int\limits^{-\vert\hat{\Omega}\vert}_{-{\sf{b}}_1}
{\mbox{d}}E\frac{E^2}
{\sqrt{-\left(E^2-{\sf{b}}_1^2\right)
\left(E^2-{\sf{b}}_2^2\right)}}
-
\frac{1}{\pi}
\int\limits_{-\vert\hat{\Omega}\vert}^{-{\sf{b}}_2}
{\mbox{d}}E\frac{\vert\hat{\Omega}\vert \vert E\vert}
{\sqrt{-\left(E^2-{\sf{b}}_1^2\right)
\left(E^2-{\sf{b}}_2^2\right)}}
\end{eqnarray}
if ${\sf{b}}_2 \le \vert\hat{\Omega}\vert<{\sf{b}}_1$, and
\begin{eqnarray}
e_0=
-\frac{1}{\pi}
\int\limits_{-{\sf{b}}_1}^{-{\sf{b}}_2}
{\mbox{d}}E\frac{E^2}
{\sqrt{-\left(E^2-{\sf{b}}_1^2\right)
\left(E^2-{\sf{b}}_2^2\right)}}
\end{eqnarray}
if $\vert\hat{\Omega}\vert<{\sf{b}}_2$.
Introducing a new variable $\varphi$ by the relation
$E=-\sqrt{{\sf{b}}_1^2-({\sf{b}}_1^2-{\sf{b}}_2^2)\sin^2\varphi}$
one gets the following final expression for the ground state energy
\begin{eqnarray}
e_0=
-\frac{1}{\pi}
\left[
{\sf{b}}_1
{\mbox{E}}
\left(
\psi,
\frac{{\sf{b}}_1^2-{\sf{b}}_2^2}{{\sf{b}}_1^2}
\right)
+
\vert\hat{\Omega}\vert\left(\frac{\pi}{2}-\psi\right)
\right]
\end{eqnarray}
where
${\mbox{E}}(\psi,a^2)=\int_0^{\psi}{\mbox{d}}\varphi
\sqrt{1-a^2\sin^2\varphi}$
is the elliptic integral of the second kind$^{50}$
and
\begin{eqnarray}
\psi=
\left\{
\begin{array}{ll}
0,
& {\mbox{if}}\;\;\;{\sf{b}}_1\le \vert\hat{\Omega}\vert,\\
{\mbox{arcsin}}
\sqrt{\frac{{\sf{b}}_1^2-\hat{\Omega}^2}
{{\sf{b}}_1^2-{\sf{b}}_2^2}},
& {\mbox{if}}\;\;\;{\sf{b}}_2\le \vert\hat{\Omega}\vert<{\sf{b}}_1,\\
\frac{\pi}{2},
& {\mbox{if}}\;\;\;\vert\hat{\Omega}\vert<{\sf{b}}_2.
\end{array}
\right.
\end{eqnarray}

The result obtained by Pincus$^{13}$
follows from (30), (31) if
$\Omega_1=\Omega_2=0$.
However,
the described approach permits to get the ground state
energy (or the Helmholtz free energy) for
more complicated regular nonuniformities
(e.g., for chains with regularly alternating non-random
or random (Lorentzian) transverse fields).
To demonstrate this let us consider at first
the spin-Peierls instability
with respect to dimerization
in the presence of a non-random transverse field.
We introduce dimerization parameter $\delta$ and assume in (30), (31)
$\vert I_1\vert=\vert I\vert(1+\delta)$,
$\vert I_2\vert=\vert I\vert(1-\delta)$,
$0\le \delta\le 1$.
Taking into account that the elastic energy per site is $\alpha\delta^2$
one must seek the minimum of the total energy 
${\cal{E}}(\delta)=e_0(\delta)+\alpha\delta^2$
as a function of $\delta$. For
${\cal{E}}(\delta)$ we find
\begin{eqnarray}
{\cal{E}}(\delta)
=-\frac{\sqrt{(\Omega_1-\Omega_2)^2+16 I^2}}
{2\pi}
{\mbox{E}}
\left(
\psi,
\frac{4 I^2(1-\delta^2)}
{\frac{1}{4}(\Omega_1-\Omega_2)^2+4 I^2}
\right)
-\vert\Omega_1+\Omega_2\vert
\left(
\frac{1}{4}-\frac{\psi}{2\pi}
\right)
+\alpha\delta^2
\end{eqnarray}
with
\begin{eqnarray}
\psi=
\left\{
\begin{array}{ll}
0,
& {\mbox{if}}\;\;\;
\sqrt{(\Omega_1-\Omega_2)^2+16 I^2}
\le \vert\Omega_1+\Omega_2\vert,\\
{\mbox{arcsin}}
\sqrt{\frac{4 I^2-\Omega_1\Omega_2}{4 I^2(1-\delta^2)}},
& {\mbox{if}}\;\;\;
\sqrt{(\Omega_1-\Omega_2)^2+16 I^2\delta^2}
\le \vert\Omega_1+\Omega_2\vert
<\sqrt{(\Omega_1-\Omega_2)^2+16 I^2},\\
\frac{\pi}{2},
& {\mbox{if}}\;\;\;
\vert\Omega_1+\Omega_2\vert
<\sqrt{(\Omega_1-\Omega_2)^2+16 I^2\delta^2} .
\end{array}
\right.
\end{eqnarray}
Eqs. (32), (33) in the limit of uniform field
$\Omega_1=\Omega_2$
coincide with the result reported in Ref. 16.
For strong fields
$\vert\Omega_1+\Omega_2\vert\ge
\sqrt{(\Omega_1-\Omega_2)^2+16 I^2}$
one finds that 
${\cal{E}}(\delta)=-\frac{1}{4}\vert\Omega_1+\Omega_2\vert+\alpha\delta^2$
and
the equation
$\frac{\partial{\cal{E}}(\delta)}{\partial\delta}=0$
has 
only the zero solution
$\delta^{\star}=0$
(no dimerization in strong enough fields),
whereas for weaker fields besides the zero solution
there may be a non-zero one
$\delta^{\star}\ne 0$
coming from the equation
\begin{eqnarray}
\alpha
=\frac{\sqrt{(\Omega_1-\Omega_2)^2+16 I^2}}
{4\pi(1-\delta^2)}
\left[
{\mbox{F}}
\left(
\psi,
\frac{4 I^2(1-\delta^2)}
{\frac{1}{4}(\Omega_1-\Omega_2)^2+4 I^2}
\right)
-
{\mbox{E}}
\left(
\psi,
\frac{4 I^2(1-\delta^2)}
{\frac{1}{4}(\Omega_1-\Omega_2)^2+4 I^2}
\right)
\right]
\end{eqnarray}
where
${\mbox{F}}(\psi,a^2)=\int_0^{\psi}{\mbox{d}}\varphi
/\sqrt{1-a^2\sin^2\varphi}$
is the elliptic integral of the first kind.$^{50}$

In the following discussion of results we choose a
uniform transverse field 
$\Omega_1=\Omega_2=\Omega_0$,
$\vert \Omega_0 \vert <2\vert I\vert$.
To give a guide for further reading this paragraph
we summarize the main results valid for sufficiently hard
lattices (having $\alpha>\frac{\vert I\vert}{4}$).
(i)
For zero field we have a minimum of the total energy
${\cal{E}}(\delta)$
at a nonzero value of the dimerization parameter
$\delta^{\star}\ne 0$.
(ii)
For finite but small fields
${\cal{E}}(\delta)$
still exhibits one minimum
at $\delta^{\star}\ne 0$
the position of which remains unchanged.
(iii)
When the field achieves a certain characteristic value
$\Omega_{0a}$
a second local minimum appears at $\delta^{\star}=0$.
The two minima at
$\delta^{\star}=0$
and
$\delta^{\star}\ne 0$
are separated by a maximum.
(iv)
At a second characteristic field
$\Omega_{0b}$
both minima at
$\delta^{\star}=0$
and
$\delta^{\star}\ne 0$
have the same depth.
(v) Further increasing $\Omega_0$
the minimum at
$\delta^{\star}=0$
becomes the global one and at a certain characteristic
field
$\Omega_{0c}$
the minimum at
$\delta^{\star}\ne 0$
abruptly disappears.
The scenario
described in (i) -- (v) is typical for a
first order transition
characterized by the order parameter $\delta^{\star}$
and driven by the transverse field $\Omega_0$.
Now we illustrate it in a more detail.

In Fig. 12 we show for different values
of $\alpha$
how the dependence of
${\cal{E}}(\delta)-{\cal{E}}(0)$
on
the dimerization parameter
varies with
the
strength of
the field
$\Omega_0$.
As it follows from Eqs. (32), (33)
(and can be also
seen in Fig. 12 where, however,
the difference
${\cal{E}}(\delta)-{\cal{E}}(0)$
is depicted)
the total energy ${\cal{E}}(\delta)$
at sufficiently large values of $\delta$
($\delta\ge\frac{\vert\Omega_0\vert}{2\vert I\vert}$,
${\cal{E}}(\delta)-{\cal{E}}(0)$ 
at the value
$\frac{\vert\Omega_0\vert}{2\vert I\vert}$
is denoted by dark circles in Fig. 12)
becomes independent
of the field.
In Fig. 13 we plot the solution of Eqs. (34), (33)
for different lattices (i.e. different values of $\alpha$)
in the presence of the field.
As a matter of fact we calculated r.h.s. of Eq. (34)
varying $\delta$ from 0 to 1 and
finding in such a way for what $\alpha$ this value of $\delta^{\star}$
realizes.
Note that solutions of Eqs. (34), (33)
$\delta^{\star}$
which are smaller than
$\frac{\vert\Omega_0\vert}{2\vert I\vert}$
realize a maximum of the total energy,
whereas solutions
$\delta^{\star}$
which are larger than
$\frac{\vert\Omega_0\vert}{2\vert I\vert}$
realize a minimum.
This can be seen, for example, for a lattice with
$\alpha=0.4$ in Figs. 12c and 13b, 13c:
at $\Omega_0=0.1$
the total energy
${\cal{E}}(\delta)$ exhibits two minima at
$\delta^{\star}=0$
and
$\delta^{\star}\ne 0$
separated by a maximum at intermediate value of
$\delta^{\star}$;
at $\Omega_0=0.2$
the total energy
${\cal{E}}(\delta)$ exhibits only a minimum at
$\delta^{\star}=0$.
From Figs. 12, 13
and Eqs. (34), (33)
one concludes
that for soft lattices having
$\alpha<\frac{\vert I\vert}{4}$
there is no solution of Eqs. (34), (33)
fulfilling the presupposition $\delta^{\star}\le 1$.
Such lattices are excluded from further consideration.
For other lattices 
the solution of Eqs. (34), (33)
$\delta^{\star}\ne 0$
existing for zero transverse field
does not feel the presence of a small field,
however, abruptly vanishes at a certain value of the
transverse field.
Moreover,
for soft lattices one needs larger fields than for hard lattices
for a disappearance of the solution of Eqs. (34), (33)
(compare Figs. 13b - 13f with Fig. 13a).
Thus, in the case of hard lattices
even small transverse fields may destroy the dimerization.
As it is seen e.g.
for
a lattice with $\alpha=0.2$  (Figs. 12, 13)
above a certain
characteristic
value of the transverse field
$\Omega_{0a}$
(for which Eqs. (34), (33) has the solution $\delta^{\star}=0$)
($\Omega_{0a}\approx 0.2$)
${\cal{E}}(\delta)$ starts to exhibit in addition to the global
minimum at $\delta^{\star}\ne 0$,
a local one at $\delta^{\star}=0$,
two minima are separated by a maximum at the intermediate value of
the dimerization parameter.
With increasing of $\Omega_0$ the depths of the minima at first become
equal
(when $\Omega_0$ has a characteristic value
$\Omega_{0b}$)
and then the minima at $\delta^{\star}=0$ becomes a global one.
The latter minima remains the only one at
$\Omega_0$ having a characteristic value
$\Omega_{0c}$
(for which Eqs. (34), (33) has the
solution $\delta^{\star}=\frac{\vert\Omega_0\vert}{2\vert I\vert}$)
($\Omega_{0c}\approx 0.5$)
that manifests a complete suppression of the dimerization by the field.
In Fig. 14 we show different regions in the plane transverse field
$\Omega_0$ -- lattice parameter $\alpha$
in which
${\cal{E}}(\delta),$ 
$0\le\delta\le 1$
exhibits one minimum at
$\delta^{\star}\ne 0$ (region A),
two minima at
$\delta^{\star}=0$ and $\delta^{\star}\ne 0$
separated by a maximum
(regions B$_1$ and B$_2$;
in the region B$_1$ the minimum at $\delta^{\star}\ne 0$ is deeper,
whereas in the region B$_2$ the minimum at $\delta^{\star}=0$ is deeper),
one minimum at
$\delta^{\star}=0$ (region C).
To find the line that separates B$_1$ and B$_2$ 
one must find for a given $\Omega_0$ such a $\delta^{\star}$ 
at which ${\cal{E}}(\delta)-{\cal{E}}(0)$ (32), (33) with 
$\alpha$ given by the r.h.s. of Eq. (34), (33) equals to zero,
and then to evaluate the r.h.s. of Eq. (34) at the sought $\delta^{\star}$.   
Crossing the phase diagram by a vertical line corresponding
to a certain lattice (e.g. with $\alpha=0.2$ in Fig. 14)
one obtains the field at which the first order transition
between the dimerized and uniform phases occur
($\Omega_{0b}$ in Fig. 14)
and the width of hysteresis
(determined by
$\Omega_{0a}$ and $\Omega_{0c}$
in Fig. 14).

Next we consider the influence of
a random Lorentzian transverse field
on the spin-Peierls instability with respect to
dimerization. For that we calculate the difference
in random-averaged total energy
(to avoid non-physical infinities
due to the Lorentzian probability distribution)
\begin{eqnarray}
\overline{{\cal{E}}(\delta)}
-\overline{{\cal{E}}(0)}
=-\frac{1}{2}\int\limits_{-\infty}^{\infty}
{\mbox{d}}E
\left(\overline{\rho_{\delta}(E)}
-\overline{\rho_0(E)}\right)
\vert E\vert
+\alpha\delta^2
\end{eqnarray}
with $\overline{\rho_{\delta}(E)}$ given by Eq. (26)
where
$\vert I_1\vert=\vert I\vert(1+\delta)$,
$\vert I_2\vert=\vert I\vert(1-\delta)$.
Let us start from the case
$\Omega_{01}=\Omega_{02}=0$,
$\Gamma_1=\Gamma_2=\Gamma$
generalizing in such a way
the consideration
for the zero transverse field by assuming the latter to be random
(Lorentzian) with the zero mean value.
As can be seen in Fig. 15
the  randomness leads
to a continuous decrease of the non-zero value of
dimerization parameter
at which
the random-averaged total energy
exhibits minimum.  At
sufficiently large strengths of disorder $\Gamma$
the minimum of
the random-averaged total energy
occurs already at
the zero
dimerization parameter, i.e.
randomness acts against dimerization and may suppress it completely for
sufficiently large strength of disorder.
Considering the equation
\begin{eqnarray}
\frac{\partial\overline{{\cal{E}}(\delta)}}{\partial\delta}
=-\frac{1}{2}
\int\limits_{-\infty}^{\infty}
{\mbox{d}}E
\frac{\partial\overline{\rho_{\delta}(E)}}{\partial\delta}
\vert E\vert
+2\alpha\delta=0
\end{eqnarray}
one can find its solution
$\delta^{\star}$ for different $\Gamma$
(see Fig. 16).
From Fig. 16 one sees
that in the case of hard lattices even small disorder may destroy the
dimerization.
In Fig. 17 we depicted different regions in the plane
strength of disorder $\Gamma$ -- lattice parameter $\alpha$
in which
$\overline{{\cal{E}}(\delta)}-\overline{{\cal{E}}(0)}$,
$0\le \delta\le 1$
exhibits one minimum at $\delta^{\star}\ne 0$
(region A) or
one minimum at $\delta^{\star}=0$
(region C). 
The boundary curve between the regions C and A is obtained by calculating 
$\alpha$ from (36) with varying $\Gamma$ for fixed $\delta=0$.
Thus, the random field with zero mean value suppresses dimerization
with increasing the strength of disorder,
however the dimerization parameter
$\delta^{\star}$ vanishes according to a second order phase transition
scenario in contrast to the previous case.

Finally we consider the case of random field with non-zero average
value, i.e.,
$\Omega_{01}=\Omega_{02}=\Omega_0\ne 0$,
$\Gamma_1=\Gamma_2=\Gamma$.
For small strengths of randomness
$\Gamma$
the above discussed scenario of one or two minimum in
$\overline{{\cal{E}}(\delta)}$
in dependence of the value of the field remains valid.
A switching on randomness for a system being in the region A
at $\Gamma=0$ (Fig. 14)
leads to continuous decreasing of $\delta^{\star}\ne 0$
to zero.
For a system being in the regions B$_1$ or B$_2$
an increasing of randomness usually leads 
at first to
a continuous decrease of
$\delta^{\star}\ne 0$ with a decrease of the depth of that minimum
and then to an abrupt disappearance of
$\delta^{\star}\ne 0$ above a certain strength of disorder. 
We also observed another influence of small randomness for a system 
being in the region B$_1$, namely, an increasing of randomness 
leads at first to a disappearance of the minimum at $\delta^{\star}=0$
that appears again for larger strength of disorder.
The details can be traced
in Fig. 18
where we plotted the dependence
$\overline{{\cal{E}}(\delta)}-\overline{{\cal{E}}(0)}$
vs $\delta$ for different $\Gamma$
considering two mean values of the random transverse field
$\Omega_0=0.1$
and
$\Omega_0=0.3$
and in Fig. 19 where we illustrated
the vanishing and appearance of the minimum at 
$\delta^{\star}=0$ with increase of randomness.
Both the one minimum profile (solid curve in Fig. 18b)
and the two minima profile (solid curves in Figs. 18c, 18e)
of that dependence existing in the non-random case
$\Gamma=0$
are finally destroyed by increasing disorder.
The phase diagrams in
the $\Gamma$ -- $\alpha$ plane
for the two mentioned values of $\Omega_0$
are shown in
Fig. 20.

Closing this Section,
we want to make some comments concerning
the conclusions on spin-Peierls
instability that can be drawn using exact results for thermodynamic
quantities of regularly nonuniform spin-$\frac{1}{2}$ isotropic $XY$
chain in a transverse field.
Although the described basic
picture of a first order phase transition in
a uniform field
seems to be qualitatively correct
we should keep in mind
that an increasing of
field at low temperature leads to a transition from dimerized to
incommensurate phase. This fact was observed experimentally and analysed
theoretically mainly for the models of CuGeO$_3$ in a number of
papers.$^{42,51-54}$
Clearly, the simple ansatz for the lattice distortion
$\delta_1\delta_2\delta_1\delta_2\ldots\;$,
$\delta_1+\delta_2=0$ permitted us to compare the ground state energies
only for dimerized and uniform phases. To detect a transition from the
dimerized to the incommensurate phase with increasing of field one may
analyse the ground state energy of a chain having larger period, say 12.
The presence of randomness requires even more complicated lattice
distortions to be examined and the continued-fraction approach for
rigorous study of thermodynamics of the regularly alternating
spin-$\frac{1}{2}$ isotropic $XY$ chain in a transverse field provides
some possibilities to perform such an analysis.
We must also keep in mind that the known spin-Peierls compounds
are described by
the spin-$\frac{1}{2}$ isotropic Heisenberg chain rather than
$XY$ chain, however, one may expect that the basic features
of the studied phenomenon
should be similar
for both quantum spin models.

\section{Summary}

To summarize, we have studied
rigorously the magnon density of states and the thermodynamics
of the periodic nonuniform
spin-$\frac{1}{2}$ isotropic $XY$ chain
in non-random/random (Lorentzian) transverse field.
We have exploited the Jordan-Wigner transformation, the temperature 
double-time Green functions and the continued fractions. The Green 
functions approach seems to be the most convenient tool for a study of 
thermodynamics of the considered spin chains since it permits to examine  
such models
with regular nonuniformity or some type of randomness or both. 
Regular
nonuniformity leads to a splitting of the magnon band into subbands
that in its turn leads to
some spectacular changes in the behaviour of 
the gap in the energy spectrum and the 
thermodynamic quantities.
In particular, 
the low-temperature dependence
of the transverse magnetization on the transverse field is composed of 
sharply
increasing parts separated by plateaus, the temperature dependence
of specific heat may exhibit a well pronounced  two-peak structure,
the temperature dependence of the initial
transverse linear susceptibility may be enhanced or suppressed.
Regularly nonuniform spin-$\frac{1}{2}$ isotropic $XY$ chain
may exhibit a non-zero transverse magnetization
at the zero average transverse field.
The regularly alternating Lorentzian disorder in the transverse field
may in specific manner influence the thermodynamic quantities leading, for
instance,
to a smearing out of only one `step'
in the step-like dependence of the
transverse magnetization versus the transverse field
at $T=0$.
The derived results for the 
(random-averaged) ground state energy permit to analyse
the effects of external non-random/random field
on the spin-Peierls instability.
Both, magnetic field as well as randomness may destroy
the dimerization
as the analysis of the (random-averaged) total energy manifests.

The presented treatment
of the regularly periodic spin-$\frac{1}{2}$ isotropic $XY$ chains
is restricted to the density of states and therefore only to thermodynamics.
It will be interesting to
study the effects of periodic nonuniformity
on spin correlations and their dynamics
especially for a model of spin-Peierls instability.
Some work for the
dynamic $zz$ spin correlations for such models has been done 
in Ref. 16.
Another interesting problem concerns
the treatment of
the periodic nonuniform spin-$\frac{1}{2}$ transverse $XY$ chains
with an
anisotropic exchange coupling
(and in particular the extremely anisotropic case, i.e.
the spin-$\frac{1}{2}$ transverse Ising chain).
Some results for 
thermodynamics of
such regularly nonuniform chains
having period 2
were obtained
in Refs. 14, 17, 23.
Their relation to the spin-Peierls instability
seems to be an intriguing issue.

\section*{Acknowledgments}

The present study was partly supported
by the DFG (projects 436 UKR 17/20/98 and Ri 615/6-1).
O. D. acknowledges the kind hospitality of
the Magdeburg University
in the spring of 1999
when this paper was completed.
The paper was discussed at the Dortmund University
and the Budapest University.
O. D. is grateful to J. Stolze
and Z. R\'{a}cz for their warm hospitality.
He also thanks to R. Lema\'{n}ski for correspondence.

\clearpage

\noindent
{\bf List of figure captions}

\vspace{1.25cm}

FIG. 1.
Magnon band structure for periodic chains
$\Omega_1I_1\Omega_2I_2\Omega_1I_1\Omega_2I_2\ldots,$
$\Omega_j=\Omega_0+\Omega_j^{\prime}$;
the shadowed areas correspond to the allowed magnon energies.
a) $\Omega_1^{\prime}+\Omega_2^{\prime}=2$,
$\vert I_1\vert=\vert I_2\vert=0.5$;
b) $\Omega_1^{\prime}+\Omega_2^{\prime}=2$,
$\vert I_1\vert=0.75$, $\vert I_2\vert=0.25$;
c) $\Omega_1^{\prime}=\Omega_2^{\prime}=1$,
$\vert I_1\vert+\vert I_2\vert=1$;
d) $\Omega_1^{\prime}=1.5$, $\Omega_2^{\prime}=0.5$,
$\vert I_1\vert+\vert I_2\vert=1$.
The horizontal lines single out the following particular chains:
$\Omega_1^{\prime}=\Omega_2^{\prime}=1$,
$\vert I_1\vert=\vert I_2\vert=0.5$ (dotted curves),
$\Omega_1^{\prime}=2$, $\Omega_2^{\prime}=0$,
$\vert I_1\vert=\vert I_2\vert=0.5$ (dashed curve),
$\Omega_1^{\prime}=\Omega_2^{\prime}=1$,
$\vert I_1\vert=0.75$, $\vert I_2\vert=0.25$ (dashed-dotted curves),
$\Omega_1^{\prime}=1.5$, $\Omega_2^{\prime}=0.5$,
$\vert I_1\vert=0.75$, $\vert I_2\vert=0.25$ (solid curves).

\vspace{1.25cm}

FIG. 2.
The density of states (a),
the dependence of the transverse magnetization
on transverse field at $T=0$ (b),
the temperature dependence of the entropy (c),
specific heat (d),
transverse magnetization (e),
and static linear transverse susceptibility (f) at $\Omega_0=0$
for
periodic chains $\Omega_1I_1\Omega_2I_2\Omega_1I_1\Omega_2I_2\ldots,$
$\Omega_j=\Omega_0+\Omega_j^{\prime}$.
The dotted curves correspond to the uniform case
$\Omega_1^{\prime}=\Omega_2^{\prime}=1$,
$\vert I_1\vert=\vert I_2\vert=0.5$,
the dashed curves correspond to the case
$\Omega_1^{\prime}=2$, $\Omega_2^{\prime}=0$,
$\vert I_1\vert=\vert I_2\vert=0.5$,
the dashed-dotted curves correspond to the case
$\Omega_1^{\prime}=\Omega_2^{\prime}=1$,
$\vert I_1\vert=0.75$, $\vert I_2\vert=0.25$,
and the solid curves correspond to the case
$\Omega_1^{\prime}=1.5$, $\Omega_2^{\prime}=0.5$,
$\vert I_1\vert=0.75$, $\vert I_2\vert=0.25$.

\vspace{1.25cm}

FIG. 3.
Magnon band structure for periodic chains
$\Omega_1I_1\Omega_2I_2\Omega_3I_3\Omega_1I_1\Omega_2I_2\Omega_3I_3\ldots,$
$\Omega_j=\Omega_0+\Omega_j^{\prime}$;
the shadowed areas correspond to the allowed magnon energies.
a) $\Omega_1^{\prime}+\Omega_2^{\prime}+\Omega_3^{\prime}=3$,
$\Omega_1^{\prime}-2\Omega_2^{\prime}+\Omega_3^{\prime}=0$,
$\vert I_1\vert=\vert I_2\vert=\vert I_3\vert=0.5$;
b) $\Omega_1^{\prime}+\Omega_2^{\prime}+\Omega_3^{\prime}=3$,
$\Omega_1^{\prime}-2\Omega_2^{\prime}+\Omega_3^{\prime}=1.5$,
$\vert I_1\vert=\vert I_2\vert=\vert I_3\vert=0.5$;
c) $\Omega_1^{\prime}+\Omega_2^{\prime}+\Omega_3^{\prime}=3$,
$\Omega_1^{\prime}-2\Omega_2^{\prime}+\Omega_3^{\prime}=0$,
$\vert I_1\vert=0.75$, $\vert I_2\vert=0.5$, $\vert I_3\vert=0.25$;
d) $\Omega_1^{\prime}+\Omega_2^{\prime}+\Omega_3^{\prime}=3$,
$\Omega_1^{\prime}-2\Omega_2^{\prime}+\Omega_3^{\prime}=1.5$,
$\vert I_1\vert=0.75$, $\vert I_2\vert=0.5$, $\vert I_3\vert=0.25$;
e) $\Omega_1^{\prime}=\Omega_2^{\prime}=\Omega_3^{\prime}=1$,
$\vert I_1\vert+\vert I_2\vert+\vert I_3\vert=1.5$,
$\vert I_1\vert-2\vert I_2\vert+\vert I_3\vert=0$;
f) $\Omega_1^{\prime}=\Omega_2^{\prime}=\Omega_3^{\prime}=1$,
$\vert I_1\vert+\vert I_2\vert+\vert I_3\vert=1.5$,
$\vert I_1\vert-2\vert I_2\vert+\vert I_3\vert=0.75$;
g) $\Omega_1^{\prime}=1.5$, $\Omega_2^{\prime}=1$, $\Omega_3^{\prime}=0.5$,
$\vert I_1\vert+\vert I_2\vert+\vert I_3\vert=1.5$,
$\vert I_1\vert-2\vert I_2\vert+\vert I_3\vert=0$;
h) $\Omega_1^{\prime}=1.5$, $\Omega_2^{\prime}=1$, $\Omega_3^{\prime}=0.5$,
$\vert I_1\vert+\vert I_2\vert+\vert I_3\vert=1.5$,
$\vert I_1\vert-2\vert I_2\vert+\vert I_3\vert=0.75$.
The horizontal lines single out the following particular chains:
$\Omega_1^{\prime}=\Omega_2^{\prime}=\Omega_3^{\prime}=1$,
$\vert I_1\vert=\vert I_2\vert=\vert I_3\vert=0.5$ (dotted curves),
$\Omega_1^{\prime}=2.5$, $\Omega_2^{\prime}=0.5$, $\Omega_3^{\prime}=0$,
$\vert I_1\vert=\vert I_2\vert=\vert I_3\vert=0.5$ (dashed curve),
$\Omega_1^{\prime}=1$, $\Omega_2^{\prime}=0.5$, $\Omega_3^{\prime}=1.5$,
$\vert I_1\vert=0.75$, $\vert I_2\vert=0.5$, $\vert I_2\vert=0.25$
(dashed-dotted curve),
$\Omega_1^{\prime}=1.5$, $\Omega_2^{\prime}=1$, $\Omega_3^{\prime}=0.5$,
$\vert I_1\vert=1$, $\vert I_2\vert=\vert I_3\vert=0.25$
(solid curve).

\vspace{1.25cm}

FIG. 4.
The same as in Fig. 2 for periodic chains
$\Omega_1I_1\Omega_2I_2\Omega_3I_3\Omega_1I_1\Omega_2I_2\Omega_3I_3\ldots,$
$\Omega_j=\Omega_0+\Omega_j^{\prime}.$
The dotted, dashed, dashed-dotted, and solid curves correspond
to the cases pointed out in the capture to Fig. 3.

\vspace{1.25cm}

FIG. 5.
The same as in Figs. 1, 3 for periodic chains
having a period 12,
$\Omega_1I_1\ldots\Omega_{12}I_{12}
\Omega_1I_1\ldots\Omega_{12}I_{12}\ldots,$
$\Omega_1=\Omega_2=\ldots=\Omega_6,$
$\Omega_7=\Omega_8=\ldots=\Omega_{12},$
$I_1=I_2=\ldots=I_6,$
$I_7=I_8=\ldots=I_{12},$
$\Omega_j=\Omega_0+\Omega_j^{\prime}$.
a) $\Omega_1^{\prime}+\Omega_7^{\prime}=2$,
$\vert I_1\vert=\vert I_7\vert=0.5$;
b) $\Omega_1^{\prime}+\Omega_7^{\prime}=2$,
$\vert I_1\vert=0.75$, $\vert I_7\vert=0.25$;
c) $\Omega_1^{\prime}=\Omega_7^{\prime}=1$,
$\vert I_1\vert+\vert I_7\vert=1$;
d) $\Omega_1^{\prime}=1.5$, $\Omega_7^{\prime}=0.5$,
$\vert I_1\vert+\vert I_7\vert=1$.
The horizontal lines single out the following particular chains:
$\Omega_1^{\prime}=\Omega_7^{\prime}=1$,
$\vert I_1\vert=\vert I_7\vert=0.5$ (dotted curves),
$\Omega_1^{\prime}=2$, $\Omega_7^{\prime}=0$,
$\vert I_1\vert=\vert I_7\vert=0.5$ (dashed curve),
$\Omega_1^{\prime}=\Omega_7^{\prime}=1$,
$\vert I_1\vert=0.75$, $\vert I_7\vert=0.25$ (dashed-dotted curves),
$\Omega_1^{\prime}=1.5$, $\Omega_7^{\prime}=0.5$,
$\vert I_1\vert=0.75$, $\vert I_7\vert=0.25$ (solid curves).

\vspace{1.25cm}

FIG. 6.
The same as in Figs. 2, 4 for the chains singled out in Fig. 5.

\vspace{1.25cm}

FIG. 7.
The dependence of the energy gap
$\Delta$
between the ground state and the first 
excited state on transverse field
$\Omega_0$
for certain regularly nonuniform
chains. a) The chain 
$\Omega_0I_1\Omega_0I_2\Omega_0I_1\Omega_0I_2\ldots\;$,
$\vert I_1\vert=0.75$,
$\vert I_2\vert=0.25$;
b) - d) the chains having periods 2, 3, and 12, respectively,
with the notations as in Figs. 2, 4, 6.

\vspace{1.25cm}

FIG. 8.
The dependence of the transverse magnetization on the transverse field
$\Omega_0$
at
$T=0$
for classical periodic nonuniform isotropic $XY$ chains in a
transverse field.
a) Chains having a period 2
($\Omega^{\prime}_1=2$, $\Omega^{\prime}_2=0$,
$\vert I_1\vert=\vert I_2\vert=0.5$ (dashed curve),
$\Omega^{\prime}_1=\Omega^{\prime}_2=1$,
$\vert I_1\vert=0.75$, $\vert I_2\vert=0.25$ (dashed-dotted curve),
$\Omega^{\prime}_1=1.5$, $\Omega^{\prime}_2=0.5$,
$\vert I_1\vert=0.75$, $\vert I_2\vert=0.25$ (solid curve));
b) chains having a period 3
($\Omega^{\prime}_1=2.5$, $\Omega^{\prime}_2=0.5$, $\Omega^{\prime}_3=0$,
$\vert I_1\vert=\vert I_2\vert=\vert I_3\vert=0.5$ (dashed curve),
$\Omega^{\prime}_1=1$, $\Omega^{\prime}_2=0.5$, $\Omega^{\prime}_3=1.5$,
$\vert I_1\vert=0.75$,
$\vert I_2\vert=0.5$,
$\vert I_3\vert=0.25$ (dashed-dotted curve),
$\Omega^{\prime}_1=1.5$, $\Omega^{\prime}_2=1$, $\Omega^{\prime}_3=0.5$,
$\vert I_1\vert=1$,
$\vert I_2\vert=\vert I_3\vert=0.25$ (solid curve));
c) chains having a period 12
($\Omega^{\prime}_1=2$, $\Omega^{\prime}_7=0$,
$\vert I_1\vert=\vert I_7\vert=0.5$ (dashed curve),
$\Omega^{\prime}_1=\Omega^{\prime}_7=1$,
$\vert I_1\vert=0.75$, $\vert I_7\vert=0.25$ (dashed-dotted curve),
$\Omega^{\prime}_1=1.5$, $\Omega^{\prime}_7=0.5$,
$\vert I_1\vert=0.75$, $\vert I_7\vert=0.25$ (solid curve)).

\vspace{1.25cm}

FIG. 9.
Illustration of the existence of a non-zero transverse magnetization
at the zero average transverse field
in a chain having period 4.
$\Omega^{\prime}_1=\Omega^{\prime}_3=0$,
$\Omega^{\prime}_2=-\Omega^{\prime}_4=-1$,
$\vert I_1\vert=\vert I_2\vert=0.5$,
$\vert I_3\vert=\vert I_4\vert=0$ (solid curves),
$\vert I_3\vert=\vert I_4\vert=0.05$ (dashed curves),
$\vert I_3\vert=\vert I_4\vert=0.25$ (dotted curves).

\vspace{1.25cm}

FIG. 10.
The random-averaged density of states (a),
the dependence of the transverse magnetization
on transverse field at $T=0$ (b),
the temperature dependence of the entropy (c),
specific heat (d),
transverse magnetization (e),
and static linear transverse susceptibility (f) at $\Omega_0=0$
for periodic chains
\linebreak
$\Omega_{01}\Gamma_1I_1\Omega_{02}\Gamma_2I_2
\Omega_{01}\Gamma_1I_1\Omega_{02}\Gamma_2I_2\ldots,$
$\Omega_{0j}=\Omega_0+\Omega_j^{\prime}$,
$\Omega_1^{\prime}=1.5$, $\Omega_2^{\prime}=0.5$,
$\vert I_1\vert=0.75$, $\vert I_2\vert=0.25$
for the case of uniform disorder $\Gamma_1=\Gamma_2=\Gamma$.
The solid curves correspond to the non-random case $\Gamma=0$;
the long-dashed curves correspond to $\Gamma=0.1$;
the short-dashed curves correspond to $\Gamma=0.25$;
the dotted curves correspond to $\Gamma=0.5$.

\vspace{1.25cm}

FIG. 11.
The same as in Fig. 10 for nonuniform disorder
$\Gamma_1\ne 0$, $\Gamma_2=0$.
The solid curves correspond to the non-random case $\Gamma_1=0$;
the long-dashed curves correspond to $\Gamma_1=0.1$;
the short-dashed curves correspond to $\Gamma_1=0.25$;
the dotted curves correspond to $\Gamma_1=0.5$.

\vspace{1.25cm}

FIG. 12.
Change of
the total energy
${\cal{E}}(\delta)-{\cal{E}}(0)$
as a function of the dimerization parameter $\delta$
in the presence of the uniform transverse field;
$\vert I\vert=0.5$;
a) $\alpha=0$,
b) $\alpha=0.2$,
c) $\alpha=0.4$;
$\Omega_0=0$ (solid curves),
$\Omega_0=0.1$ (dashed-dotted-dotted curves),
$\Omega_0=0.2$ (dashed-dotted curves),
$\Omega_0=0.3$ (dashed curves),
$\Omega_0=0.4$ (dotted curves).

\vspace{1.25cm}

FIG. 13.
Dimerization parameter $\delta^{\star}$
as a function of $\alpha$
in the presence of a uniform transverse field $\Omega_0$;
$\vert I\vert=0.5$;
$\Omega_0=0$ (a),
$\Omega_0=0.1$ (b),
$\Omega_0=0.2$ (c),
$\Omega_0=0.3$ (d),
$\Omega_0=0.4$ (e),
$\Omega_0=0.5$ (f).
The solid curves show the solution of Eqs. (34), (33) corresponding to a
minimum of the total energy;
the dashed curve in (a) corresponds to the 
dependence $\delta^{\star}$ versus $\alpha$
valid for hard lattices 
that was
obtained in Ref. 13;
the dashed curves in (b) - (f) 
show the solution of Eqs. (34), (33)
corresponding to a maximum
of the total energy.

\vspace{1.25cm}

FIG. 14.
Different types of solution for
the dimerization parameter $\delta^{\star}$
($0\le \delta^{\star}\le 1$)
in the plane
$\Omega_0$ -- $\alpha$;
$\vert I\vert=0.5$.
Region A:
${\cal{E}}(\delta)$ has one minimum at
$\delta^{\star}\ne 0$,
regions B$_1$, B$_2$:
${\cal{E}}(\delta)$ has two minima at
$\delta^{\star}=0$ (favourable in B$_2$)
and
$\delta^{\star}\ne 0$ (favourable in B$_1$)
separated by
a maximum,
moreover,
the depths of the minima
at the line
that separates B$_1$ and B$_2$
are the same;
region C:
${\cal{E}}(\delta)$ has one minimum at
$\delta^{\star}=0$.

\vspace{1.25cm}

FIG. 15.
Change of the
random-averaged total energy
as a function of the dimerization parameter
in the presence of 
a uniform random Lorentzian transverse field with zero mean
value;
$\vert I\vert=0.5$,
$\Gamma_1=\Gamma_2=\Gamma=0$ (solid curves),
$\Gamma=0.02$ (dashed-dotted curves),
$\Gamma=0.1$ (dashed curves),
$\Gamma=0.5$ (dotted curves);
a) $\alpha=0$,
b) $\alpha=0.2$,
c) $\alpha=0.4$.

\vspace{1.25cm}

FIG. 16.
The solution of Eq. (36) as a function of $\alpha$
in the presence of disorder;
$\vert I\vert=0.5$,
$\Omega_{01}=\Omega_{02}=0$,
$\Gamma_1=\Gamma_2=\Gamma=0$ (solid curves),
$\Gamma=0.02$ (dashed-dotted curves),
$\Gamma=0.1$ (dashed curves),
$\Gamma=0.5$ (dotted curves).

\vspace{1.25cm}

FIG. 17.
Different types of solution for the dimerization parameter $\delta^{\star}$
in the plane $\Gamma$ -- $\alpha$;
$\vert I\vert=0.5$,
$\Omega_0=0$.
Region A:
$\overline{{\cal{E}}(\delta)}
-\overline{{\cal{E}}(0)}$
has one minimum at $\delta^{\star}\ne 0$,
region C:
$\overline{{\cal{E}}(\delta)}
-\overline{{\cal{E}}(0)}$
has one minimum at $\delta^{\star}=0$.

\vspace{1.25cm}

FIG. 18.
Change of the
random-averaged total energy
as a function of the dimerization parameter
in the presence of 
the uniform random Lorentzian transverse field with a non-zero mean
value
$\Omega_0=0.1$ (a, b, c)
and
$\Omega_0=0.3$ (d, e, f);
$\vert I\vert=0.5$,
$\Gamma_1=\Gamma_2=\Gamma=0$ (solid curves),
$\Gamma=0.02$ (dashed-dotted curves),
$\Gamma=0.1$ (dashed curves),
$\Gamma=0.5$ (dotted curves);
$\alpha=0$ (a, d),
$\alpha=0.2$ (b, e),
$\alpha=0.4$ (c, f).

\vspace{1.25cm}

FIG. 19.
Change of $\overline{{\cal{E}}(\delta)}-\overline{{\cal{E}}(0)}$ 
as a function of $\delta$ in the presence of the uniform random  Lorentzian 
transverse field with 
$\Omega_0=0.3$, 
$\Gamma=0.01$ (solid curves),
$\Gamma=0.1$ (dashed-dotted curves),
$\Gamma=0.2$ (dashed curves),
$\Gamma=0.3$ (dotted curves);
$\vert I\vert=0.5$, $\alpha=0.15$.

\vspace{1.25cm}

FIG. 20.
Different types of solution for the dimerization parameter
$\delta^{\star}$ in the plane
in the plane $\Gamma$ -- $\alpha$;
$\vert I\vert=0.5$,
$\Omega_0=0.1$ (a),
$\Omega_0=0.3$ (b).
Region A:
$\overline{{\cal{E}}(\delta)}
-\overline{{\cal{E}}(0)}$
has one minimum at $\delta^{\star}\ne 0$,
region B$_1$:
$\overline{{\cal{E}}(\delta)}
-\overline{{\cal{E}}(0)}$
has two minima at
$\delta\ne 0$
and
$\delta=0$
and the first one is favourable,
region B$_2$:
$\overline{{\cal{E}}(\delta)}
-\overline{{\cal{E}}(0)}$
has two minima at
$\delta\ne 0$
and
$\delta=0$
and the second one is favourable,
region C:
$\overline{{\cal{E}}(\delta)}
-\overline{{\cal{E}}(0)}$
has one minimum at $\delta^{\star}=0$.


\clearpage

\begin{figure}
\epsfxsize=120mm
\epsfbox{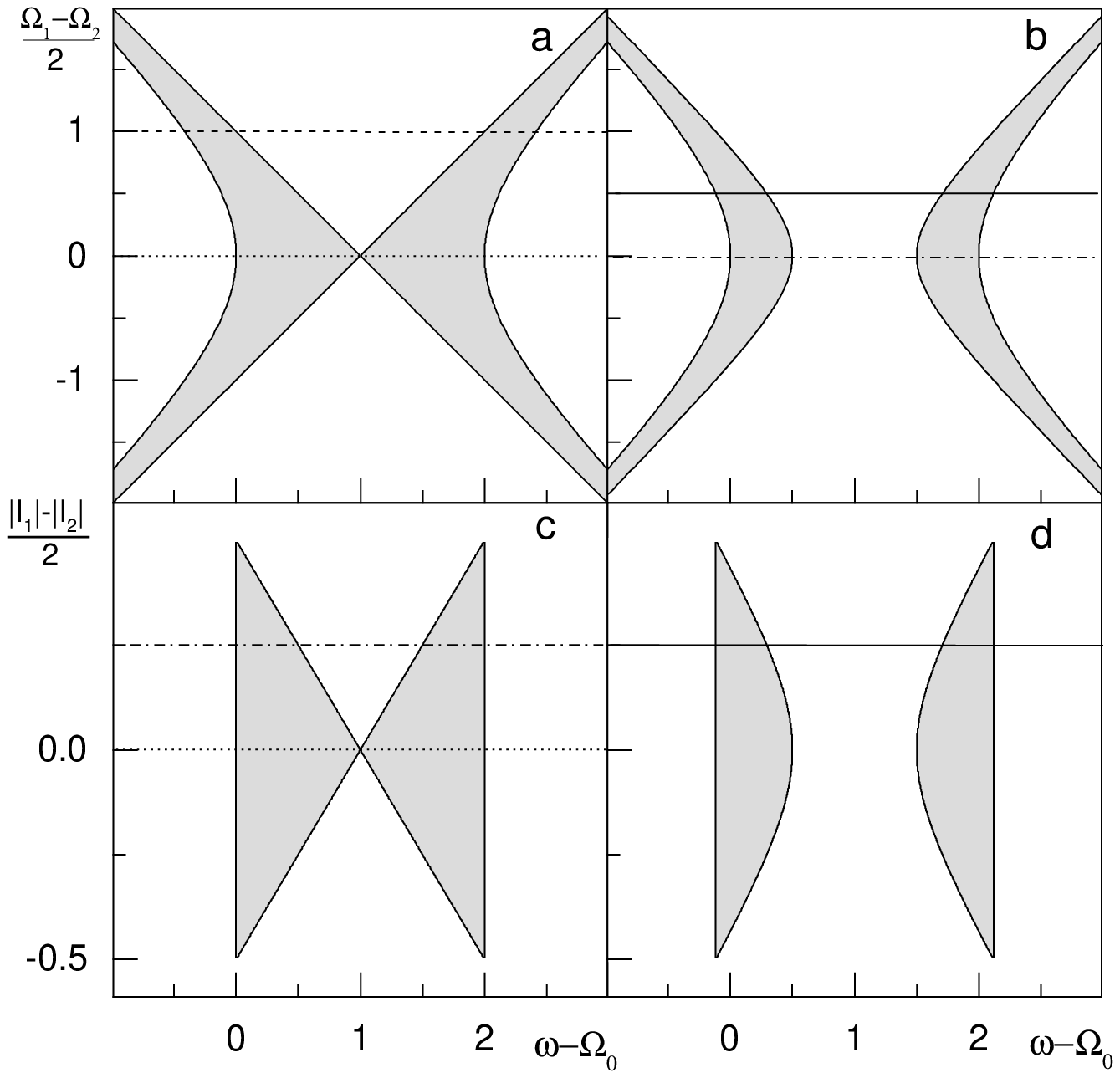}
\vspace{15mm}
\caption{FIGURE 1.}
\end{figure}

\clearpage

\begin{figure}
\epsfxsize=120mm
\epsfbox{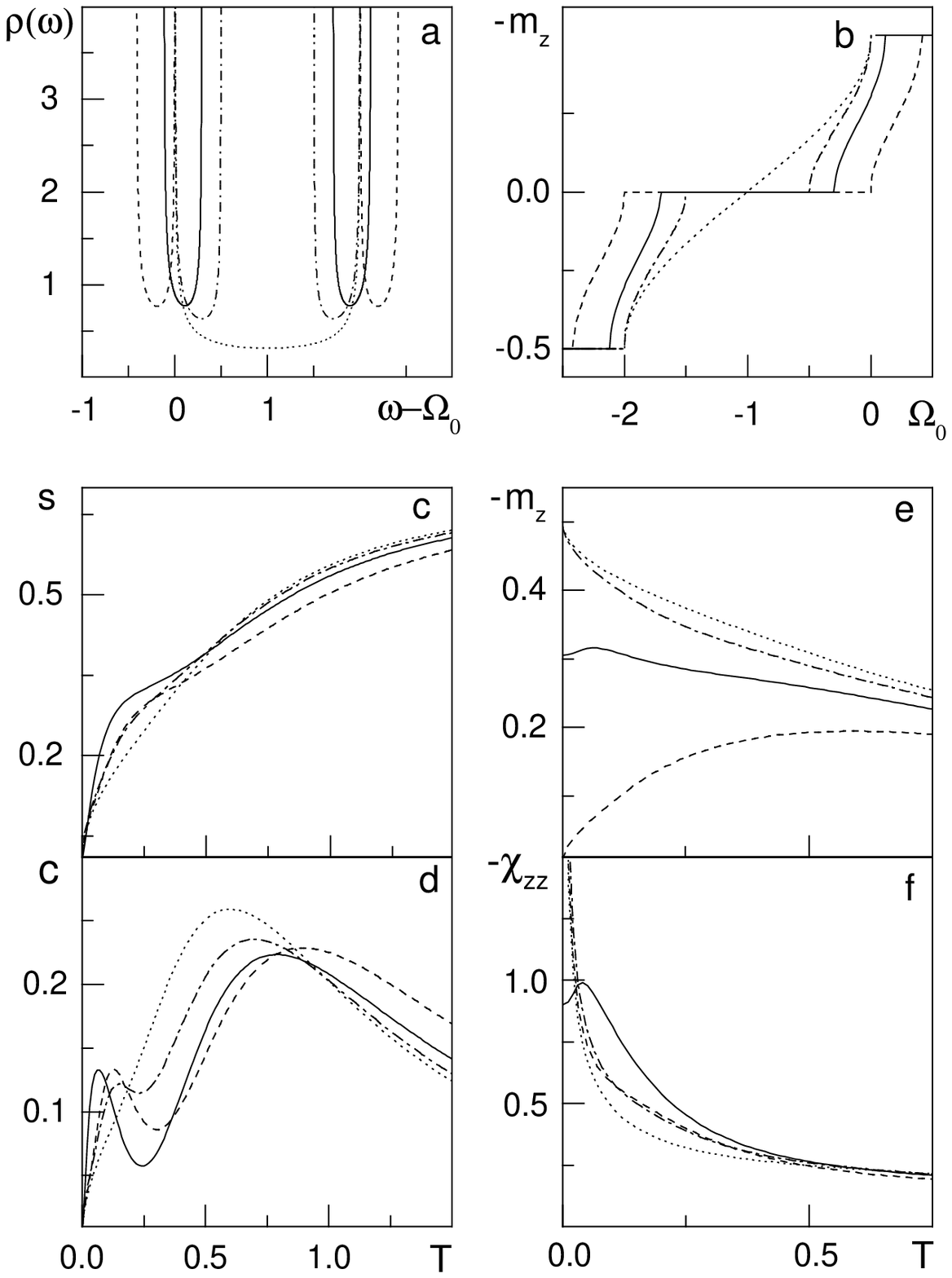}
\vspace{15mm}
\caption{FIGURE 2.}
\end{figure}

\clearpage

\begin{figure}
\epsfxsize=120mm
\epsfbox{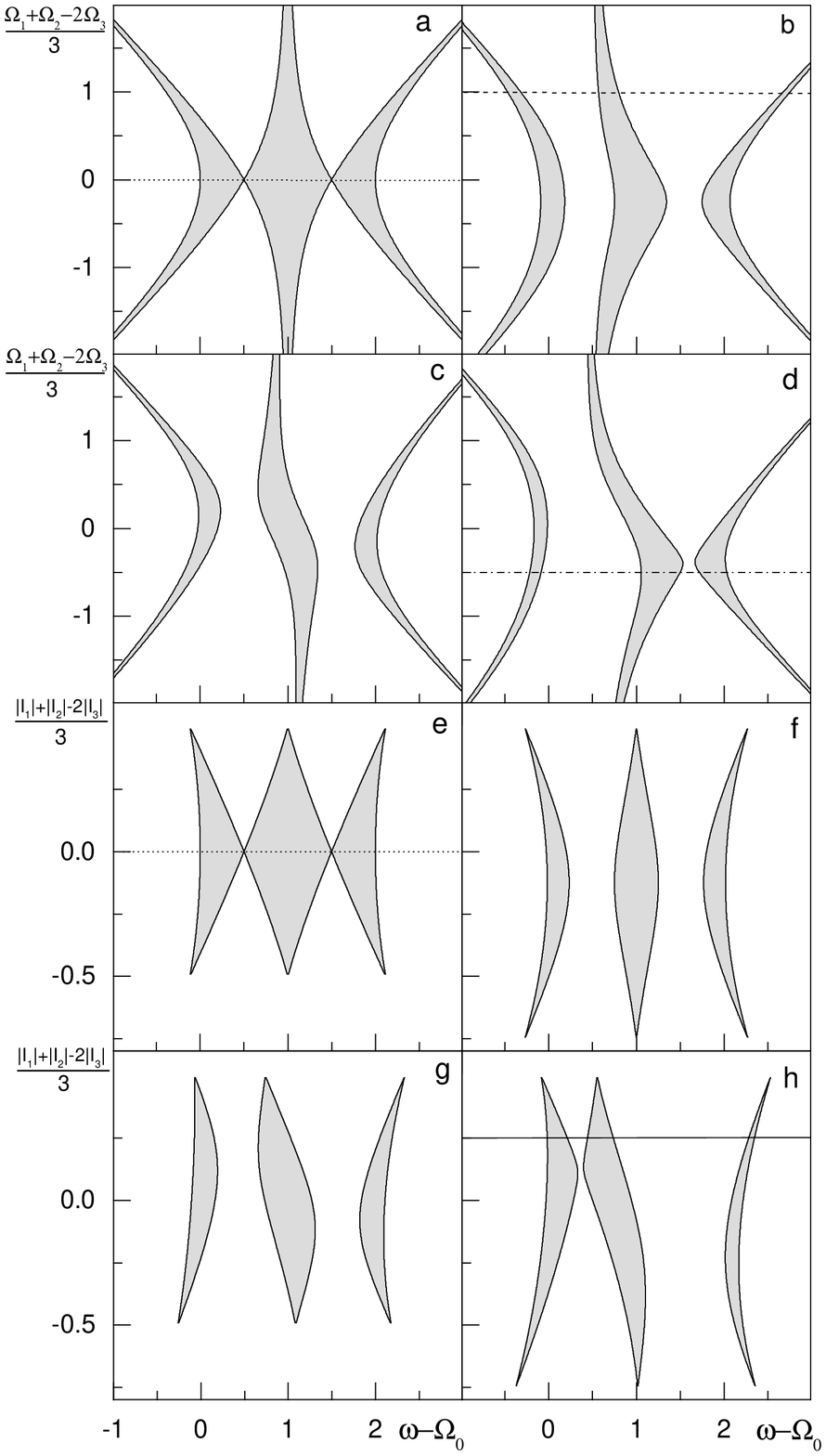}
\vspace{15mm}
\caption{FIGURE 3.}
\end{figure}

\clearpage

\begin{figure}
\epsfxsize=120mm
\epsfbox{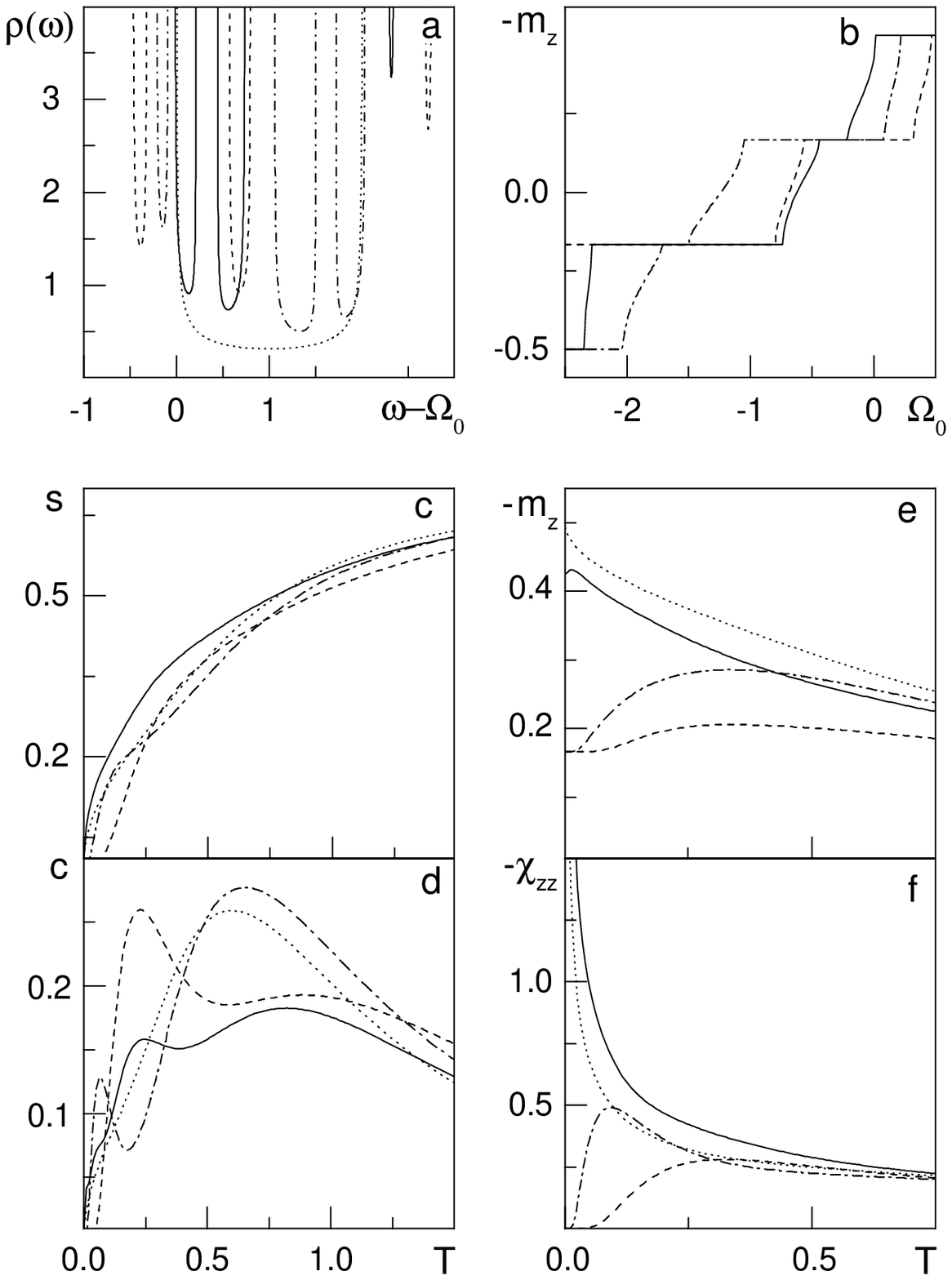}
\vspace{15mm}
\caption{FIGURE 4.}
\end{figure}

\clearpage

\begin{figure}
\epsfxsize=120mm
\epsfbox{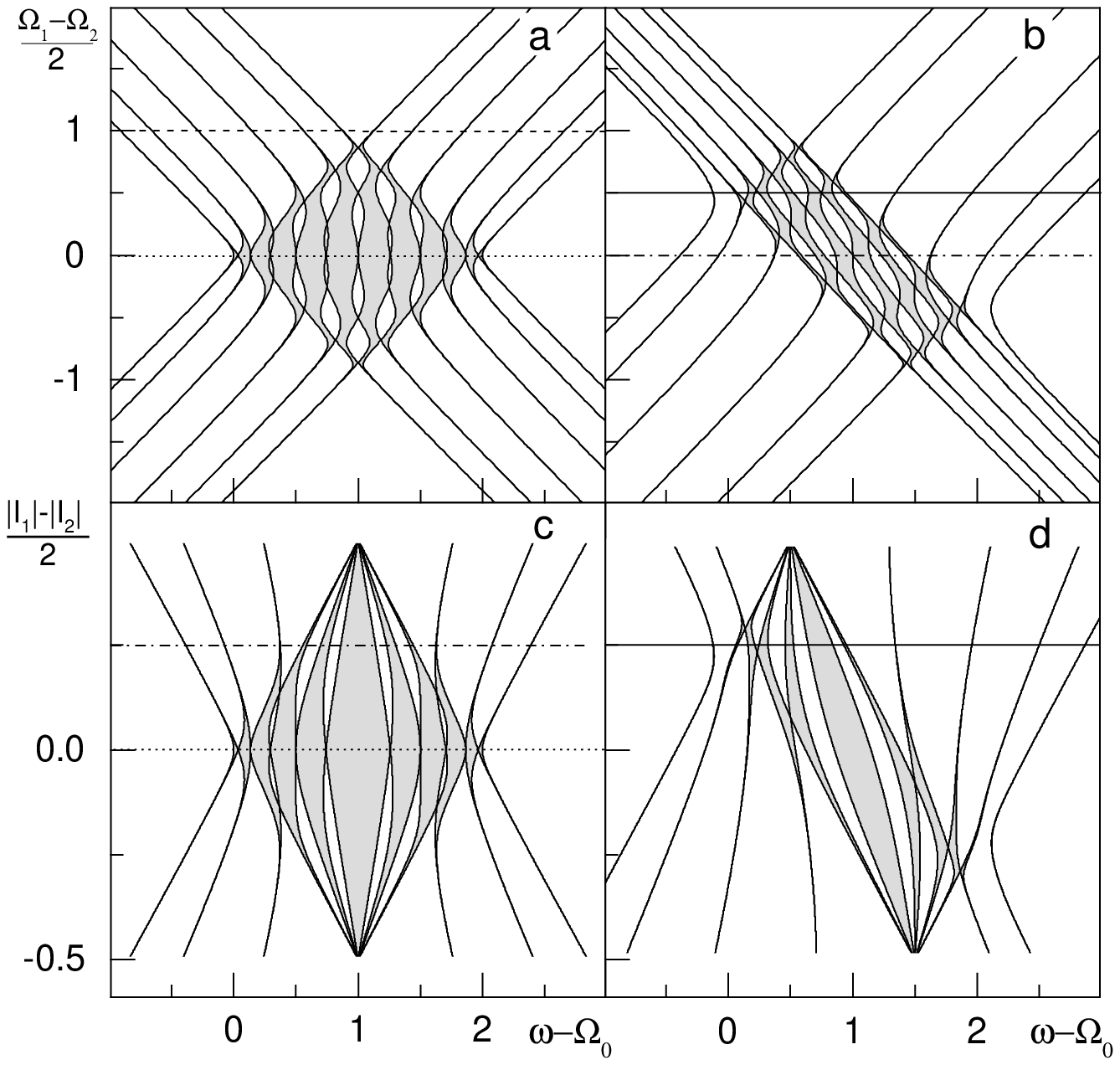}
\vspace{15mm}
\caption{FIGURE 5.}
\end{figure}

\clearpage

\begin{figure}
\epsfxsize=120mm
\epsfbox{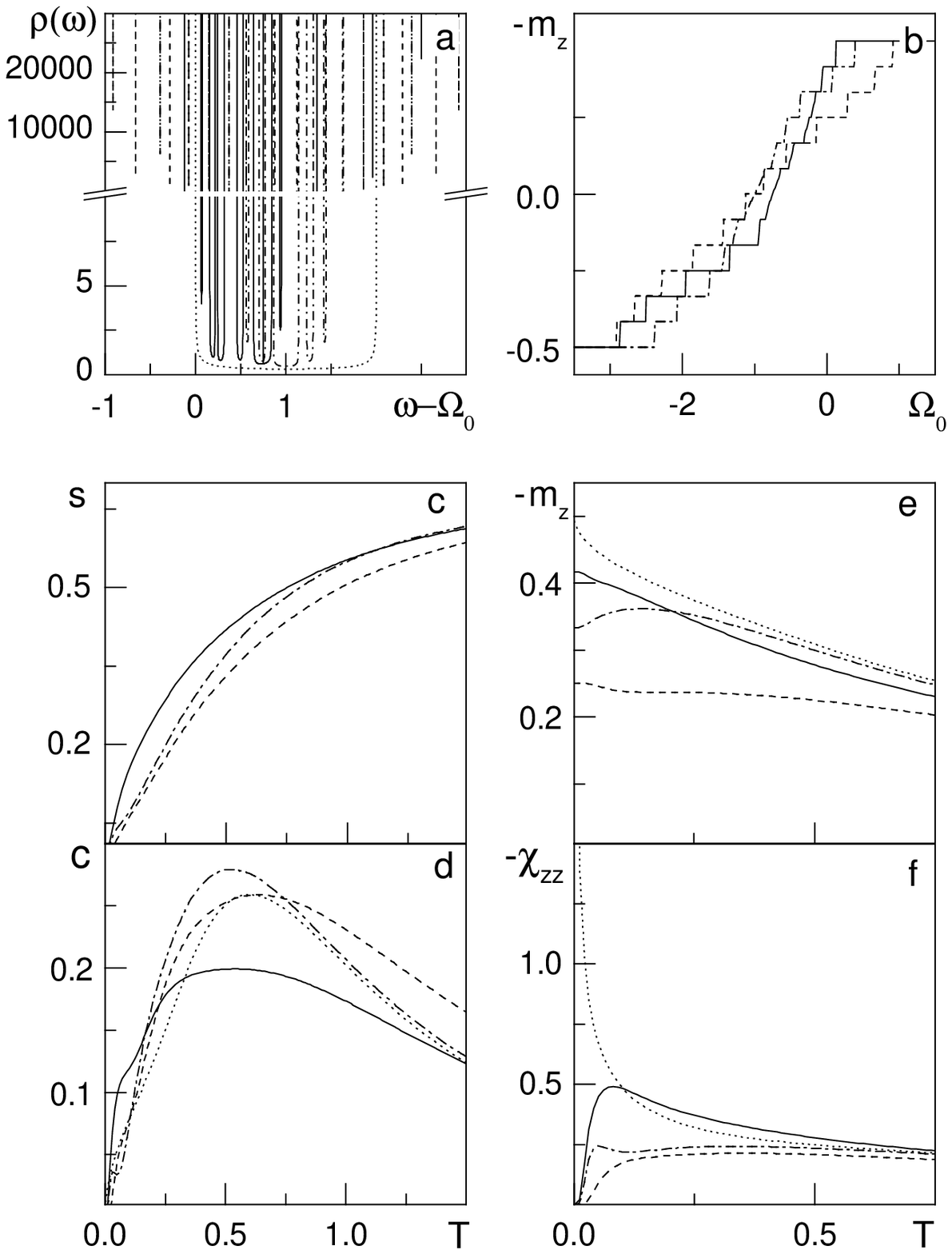}
\vspace{15mm}
\caption{FIGURE 6.}
\end{figure}

\clearpage

\begin{figure}
\epsfxsize=120mm
\epsfbox{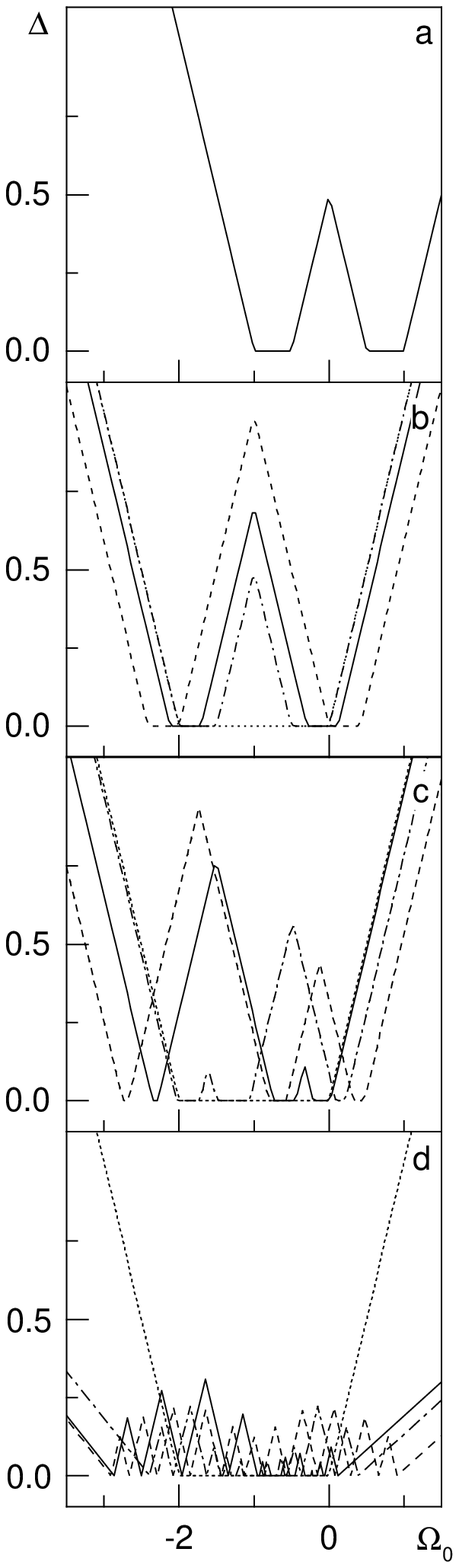}
\vspace{15mm}
\caption{FIGURE 7.}
\end{figure}

\clearpage

\begin{figure}
\epsfxsize=120mm
\epsfbox{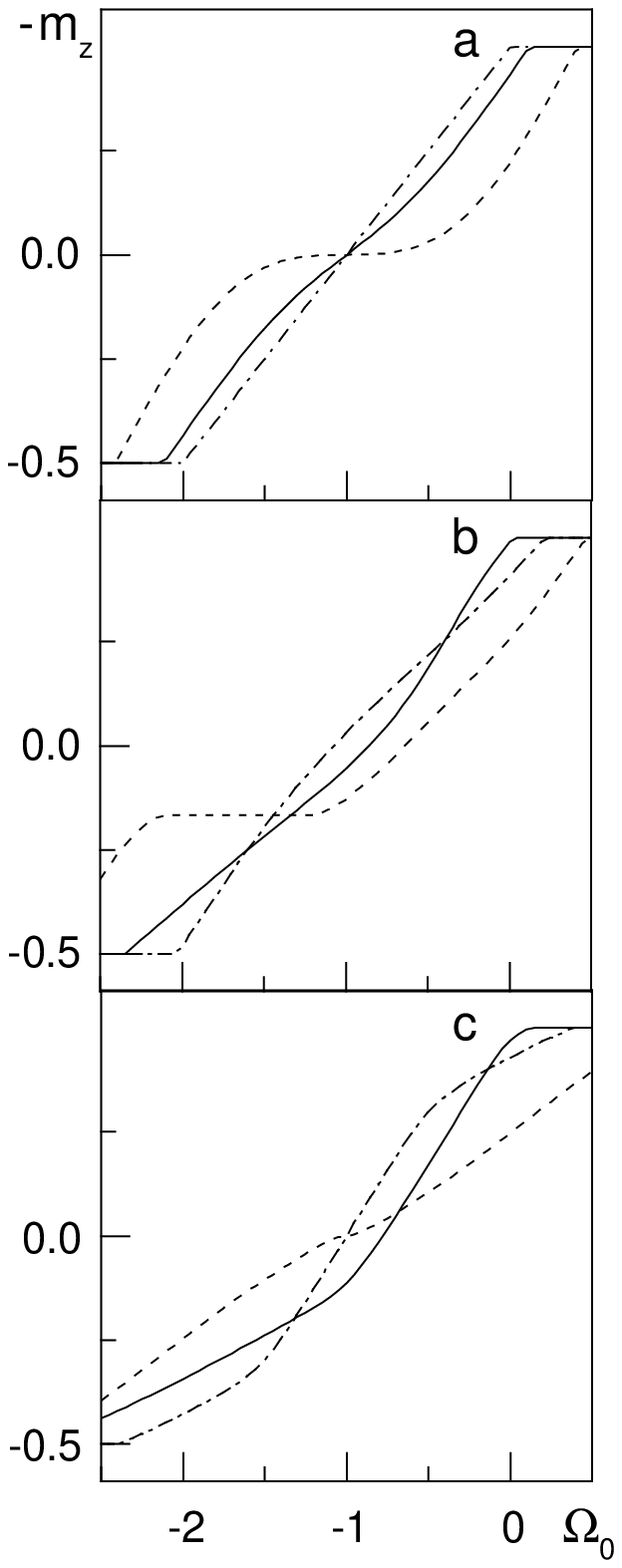}
\vspace{15mm}
\caption{FIGURE 8.}
\end{figure}

\clearpage

\begin{figure}
\epsfxsize=120mm
\epsfbox{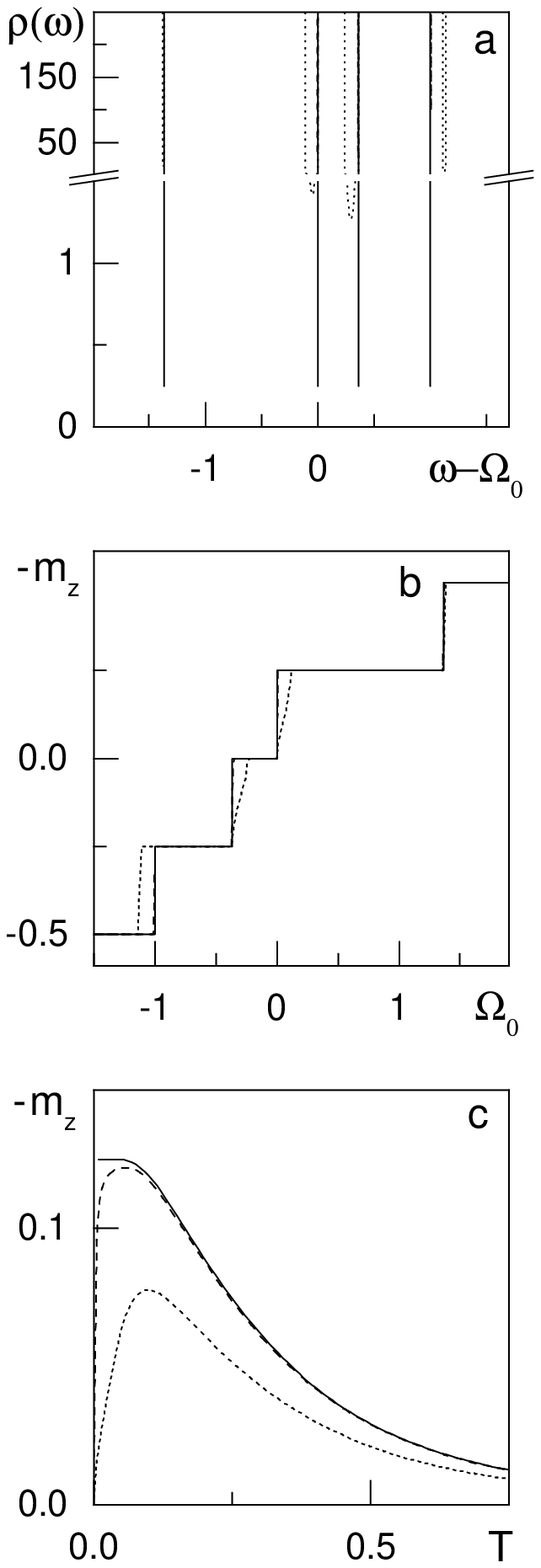}
\vspace{15mm}
\caption{FIGURE 9.}
\end{figure}

\clearpage

\begin{figure}
\epsfxsize=120mm
\epsfbox{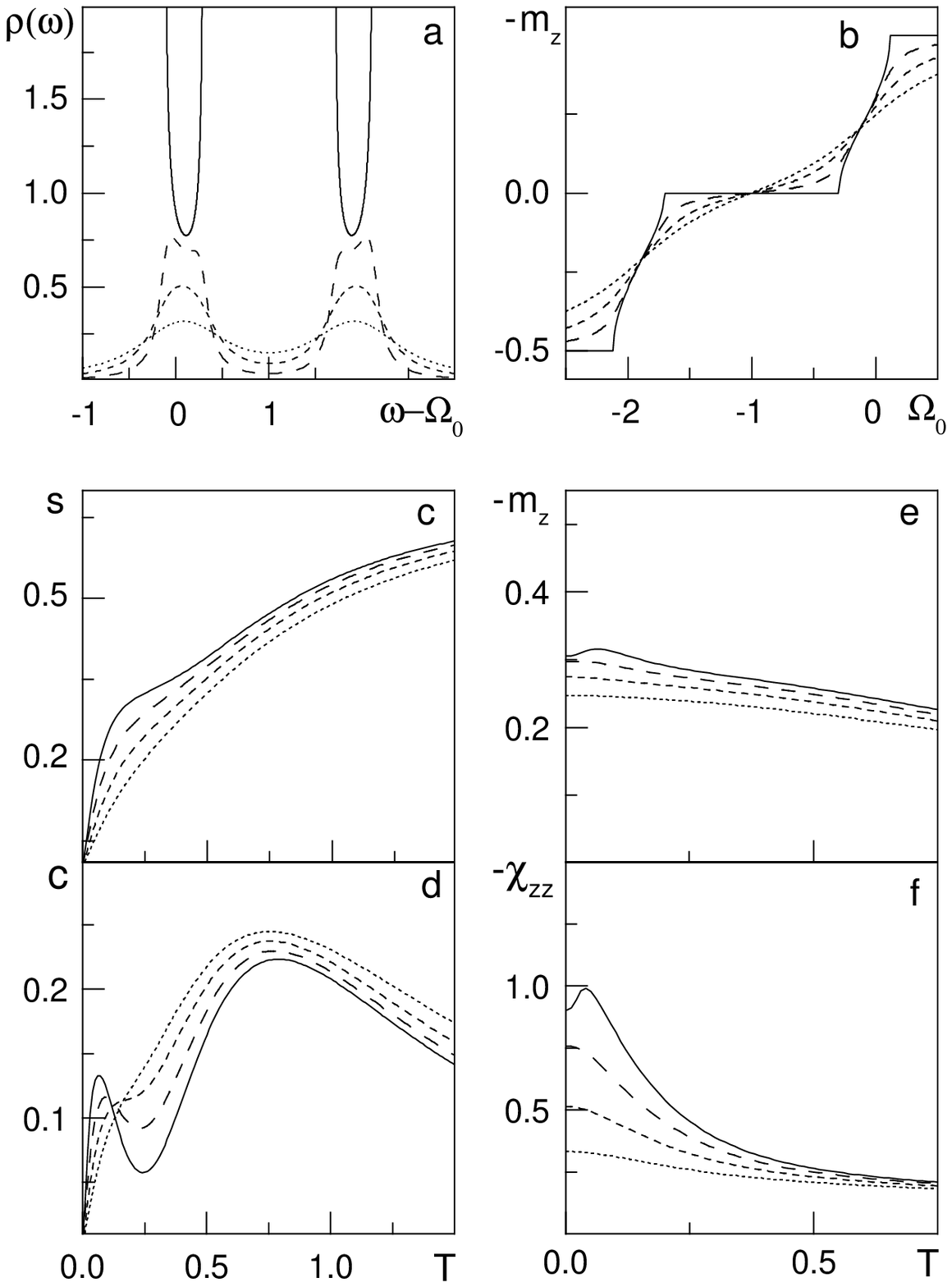}
\vspace{15mm}
\caption{FIGURE 10.}
\end{figure}

\clearpage

\begin{figure}
\epsfxsize=120mm
\epsfbox{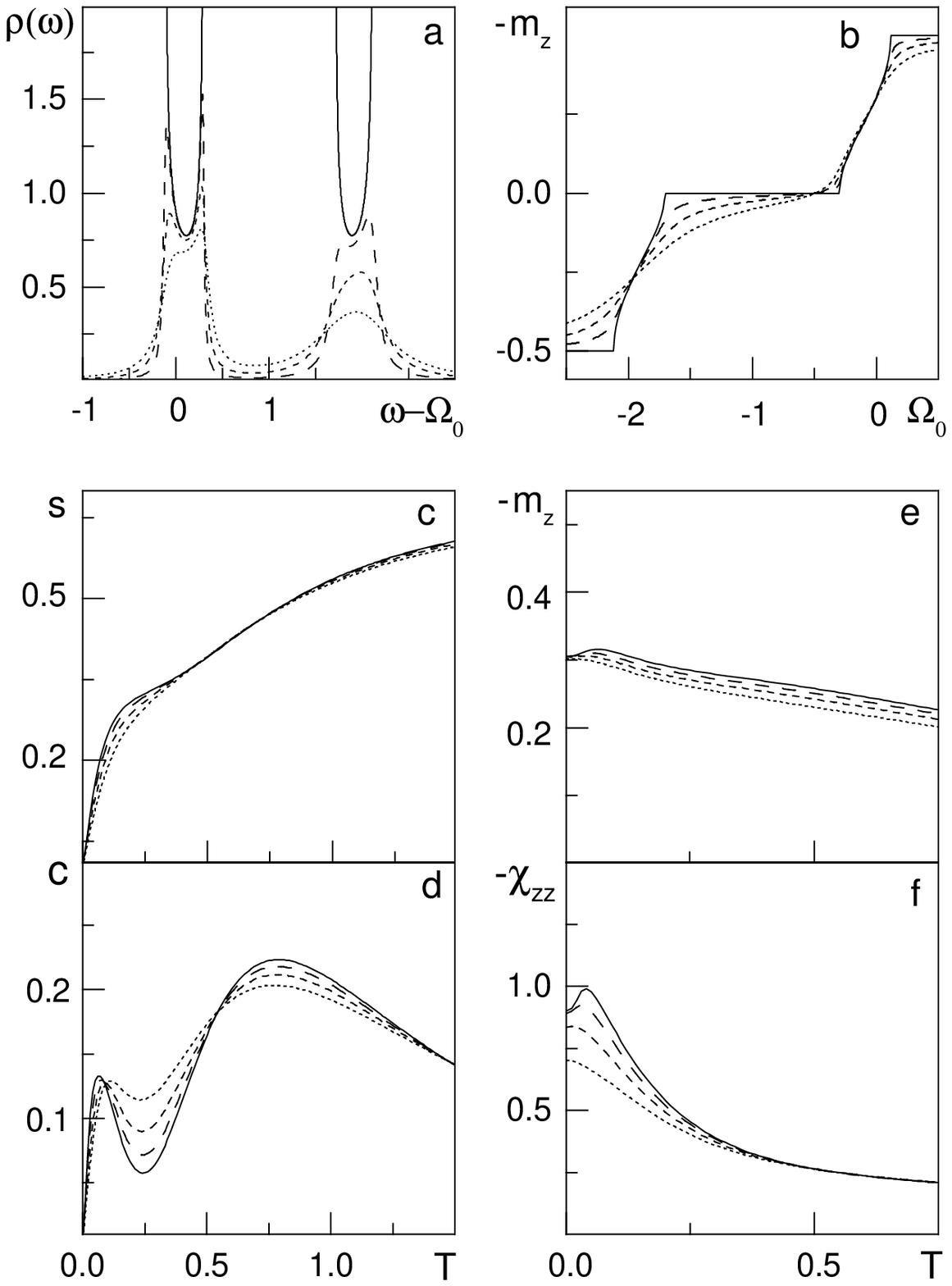}
\vspace{15mm}
\caption{FIGURE 11.}
\end{figure}

\clearpage

\begin{figure}
\epsfxsize=150mm
\epsfbox{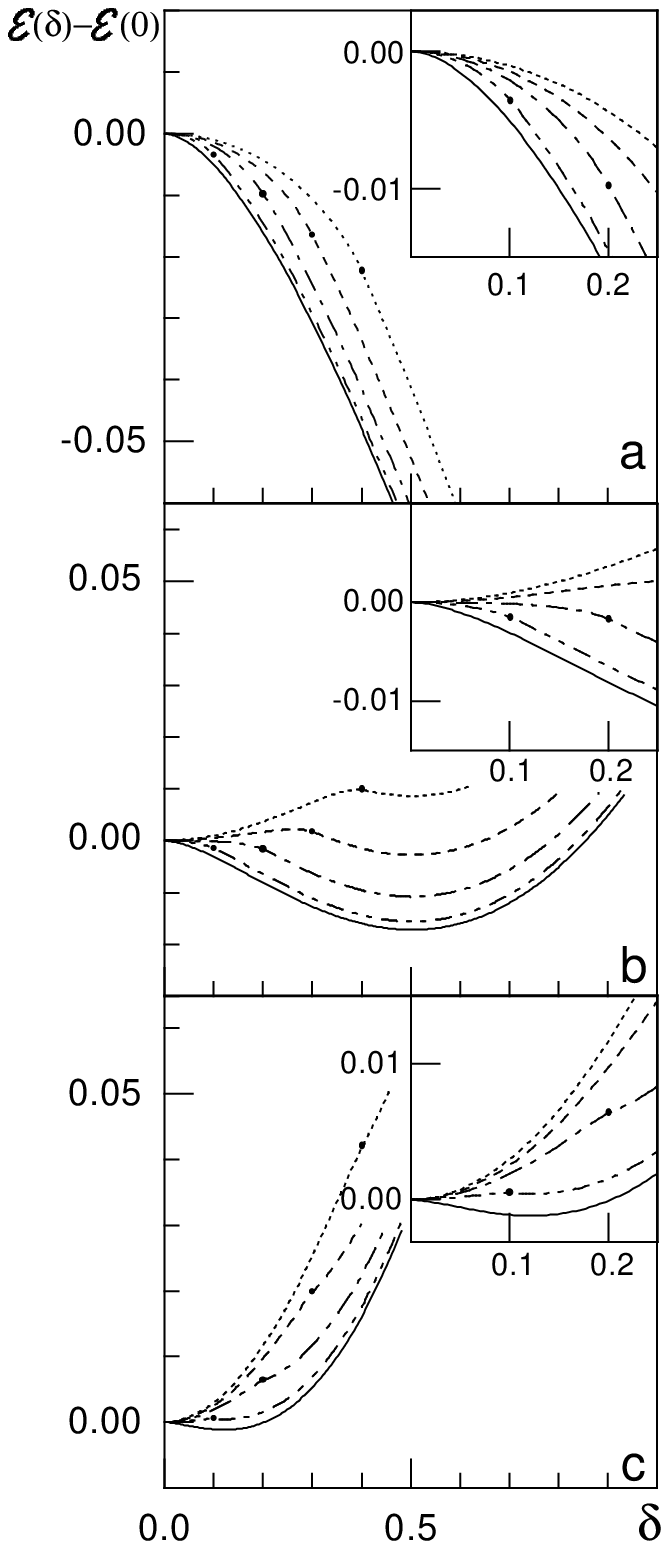}
\vspace{-30mm}
\caption{FIGURE 12.}
\end{figure}

\clearpage

\begin{figure}
\epsfxsize=140mm
\epsfbox{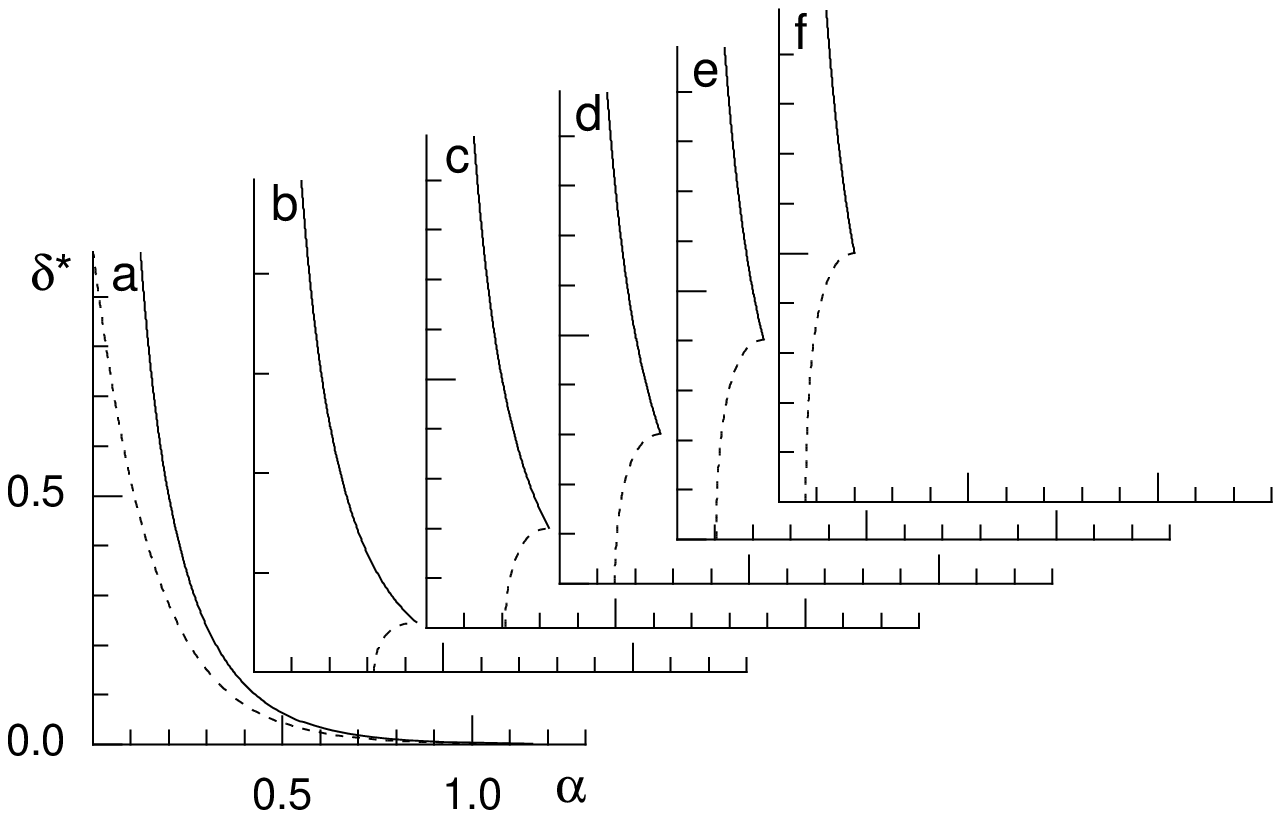}
\vspace{15mm}
\caption{FIGURE 13.}
\end{figure}

\clearpage

\begin{figure}
\epsfxsize=200mm
\epsfbox{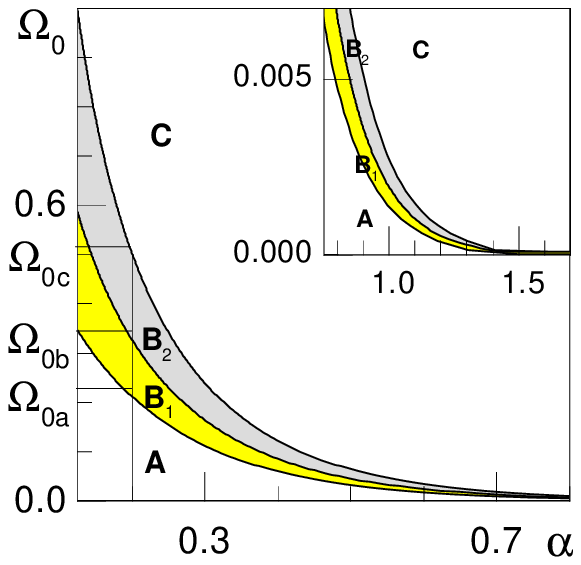}
\vspace{-60mm}
\caption{FIGURE 14.}
\end{figure}

\clearpage

\begin{figure}
\epsfxsize=150mm
\epsfbox{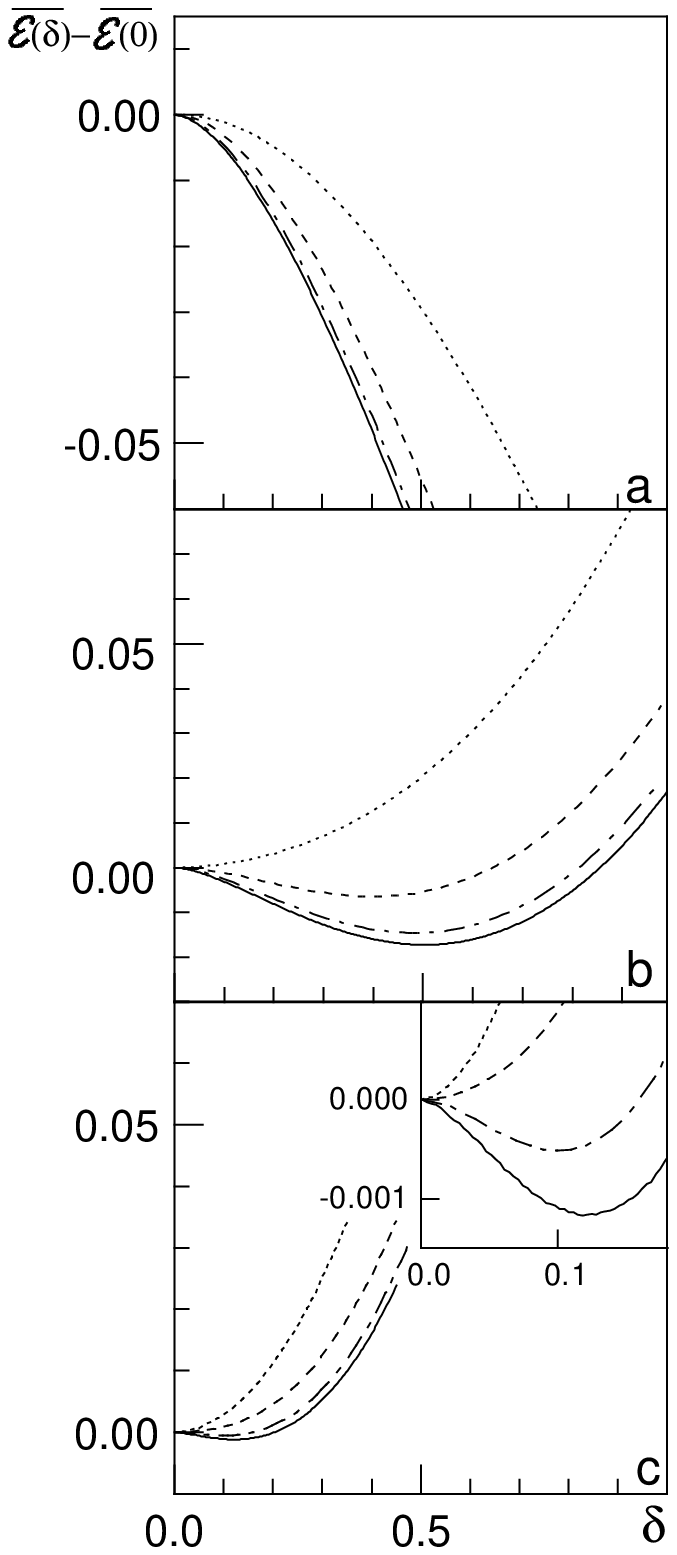}
\vspace{-30mm}
\caption{FIGURE 15.}
\end{figure}

\clearpage

\begin{figure}
\vspace{15mm}
\epsfxsize=180mm
\epsfbox{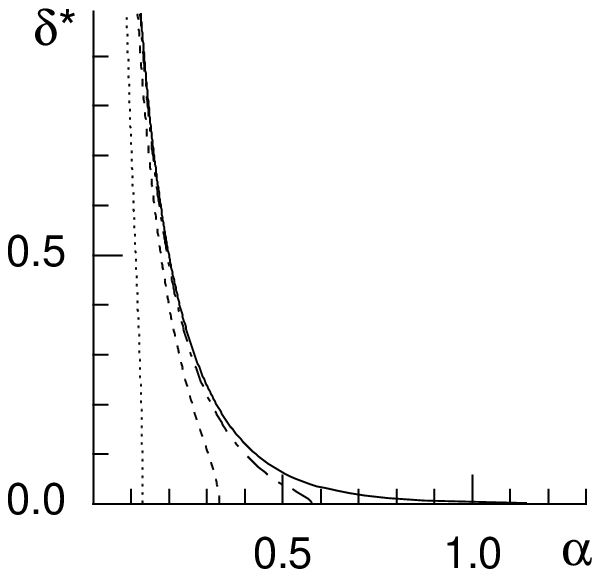}
\vspace{-35mm}
\caption{FIGURE 16.}
\end{figure}

\clearpage

\begin{figure}
\epsfxsize=200mm
\epsfbox{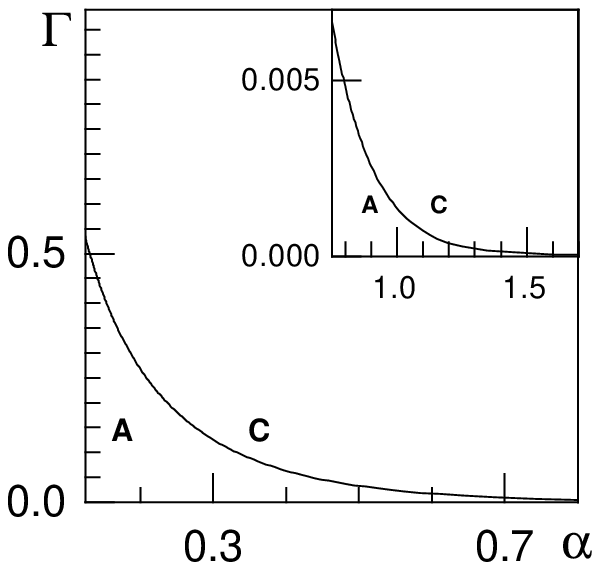}
\vspace{-60mm}
\caption{FIGURE 17.}
\end{figure}

\clearpage

\begin{figure}
\epsfxsize=150mm
\epsfbox{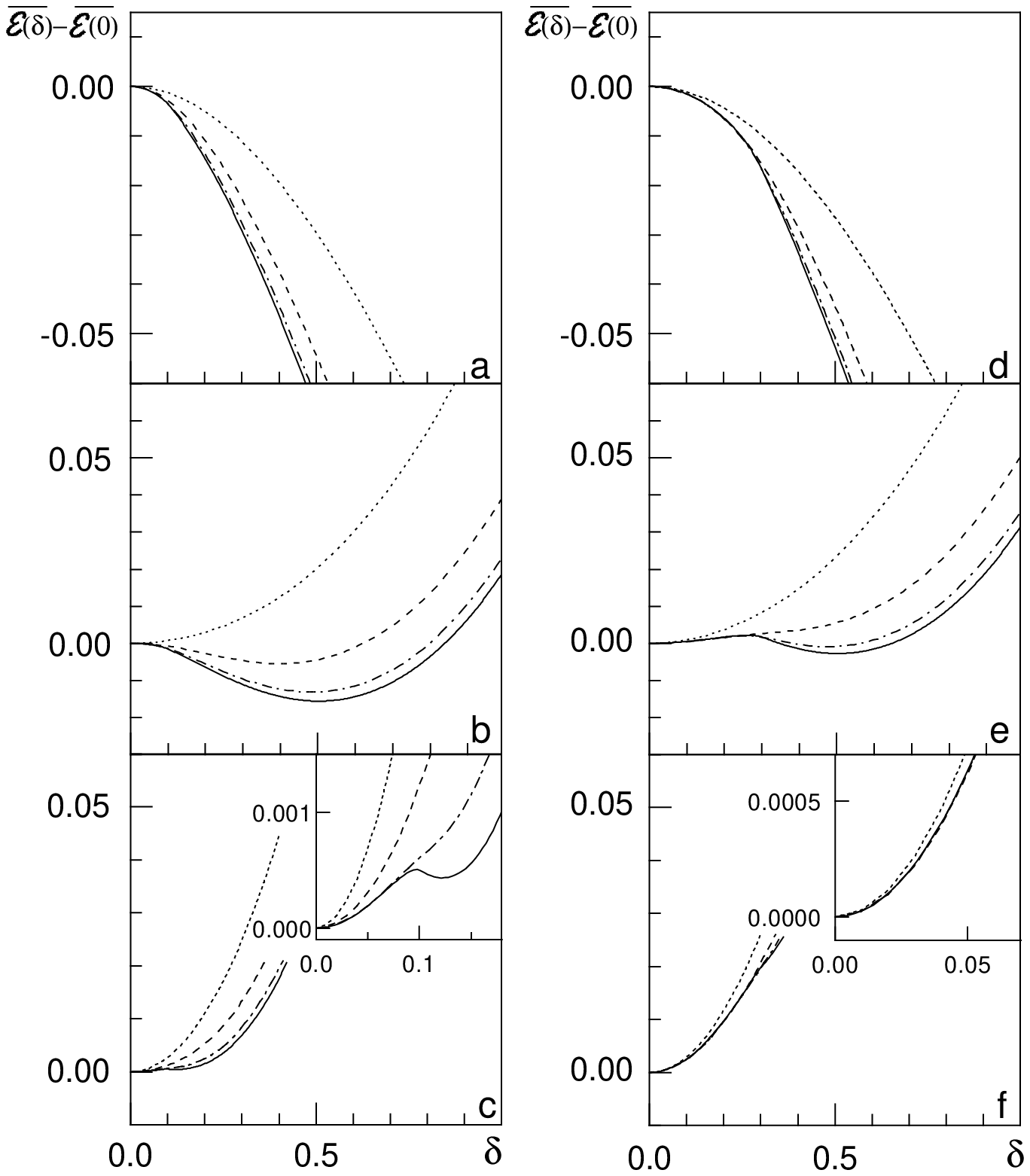}
\vspace{-30mm}
\caption{FIGURE 18.}
\end{figure}

\clearpage

\begin{figure}
\epsfxsize=150mm
\epsfbox{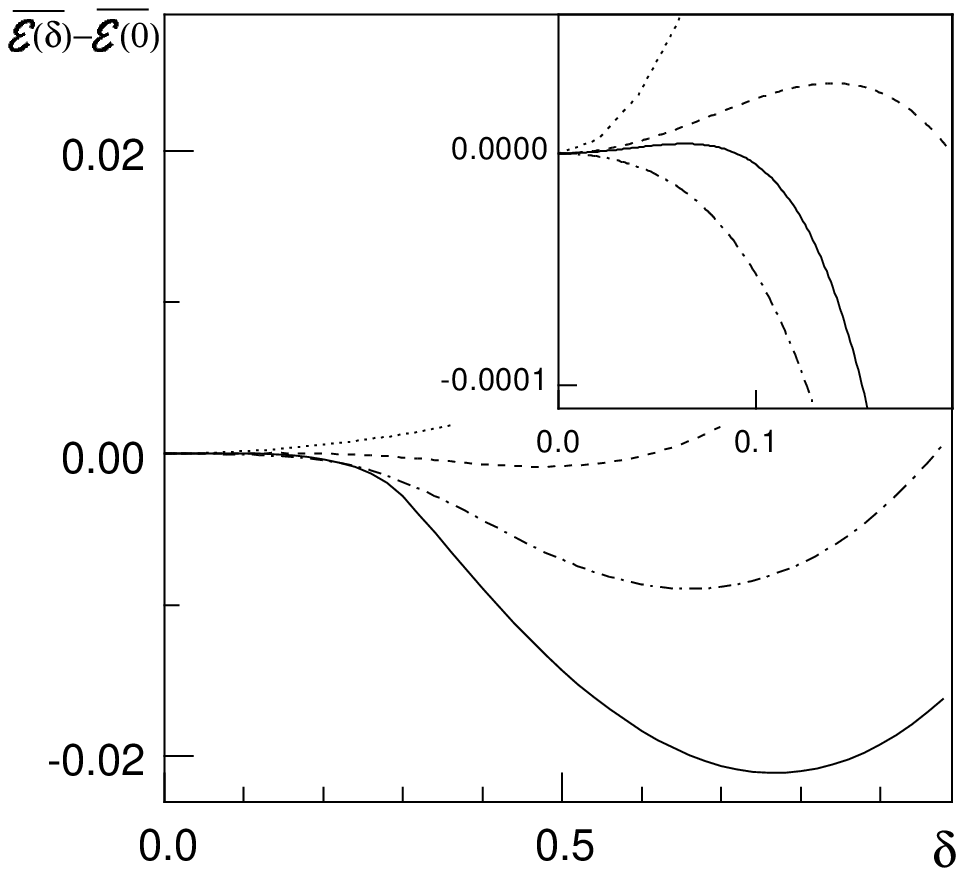}
\vspace{-60mm}
\caption{FIGURE 19.}
\end{figure}

\clearpage

\begin{figure}
\epsfxsize=180mm
\epsfbox{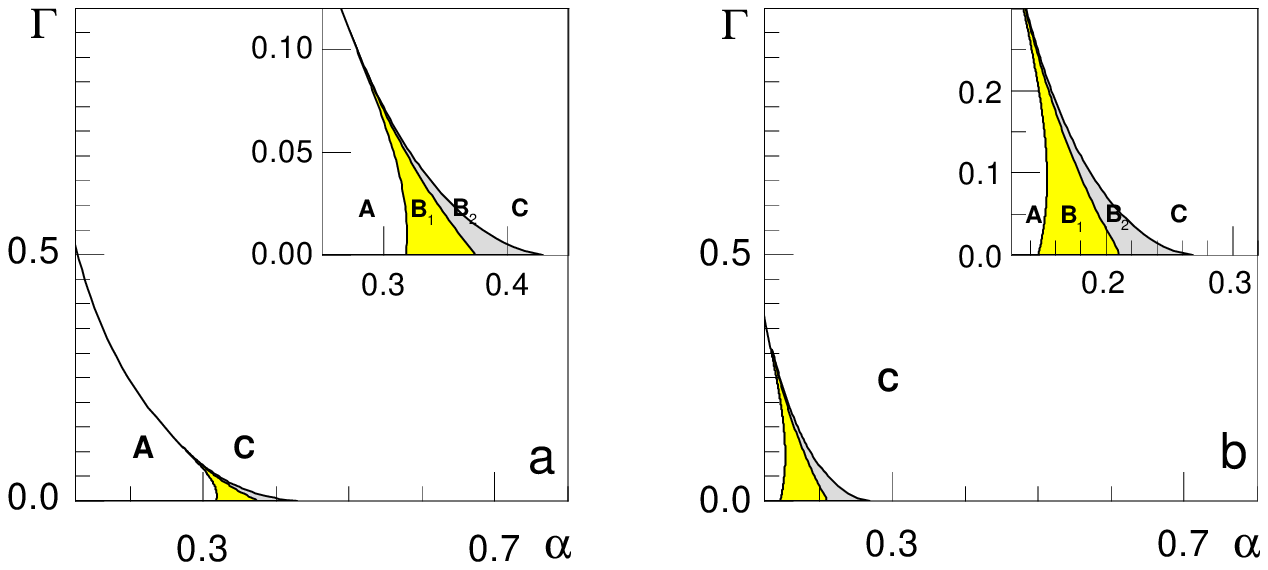}
\vspace{-60mm}
\caption{FIGURE 20.}
\end{figure}